\documentclass[12pt,preprint]{aastex}

\slugcomment{Submitted to ApJ}

\shorttitle{SMALL STRUCTURES VIA THERMAL INSTABILITY OF PARTIALLY IONIZED PLASMA I.}
\shortauthors{Fukue and Kamaya}

\begin{document}



\title{SMALL STRUCTURES VIA THERMAL INSTABILITY OF PARTIALLY IONIZED PLASMA I. CONDENSATION MODE}



\author
{TSUBASA FUKUE 
 AND HIDEYUKI KAMAYA \altaffilmark{1}
}
\affil{Department of Astronomy, School of Science, Kyoto University,
    Kyoto 606-8502, Japan}

\altaffiltext{1}{Current address: Department of Earth and Ocean Sciences, School of Applied Sciences, 
National Defense Academy of Japan, 
Yokosuka 239-8686, Japan}

\email{tsubasa@kusastro.kyoto-u.ac.jp}




\begin{abstract}

Thermal instability of partially ionized plasma is investigated by
means of a linear perturbation analysis. According to the previous
studies under the one fluid approach, the thermal instability
is suppressed due to the magnetic pressure. However, the previous
studies did not precisely consider the effect of the ion-neutral
friction, since they did not treat the flow as two fluid which is
composed of ions and neutrals. Then, we revisit the effect of the
ion-neutral friction of the two fluid
to the growth of the thermal instability. According to our study,
the characteristic features of the instability are the following four
points:
(1) The instability which is characterized by the mean molecular
weight of neutrals
is suppressed via the ion-neutral friction only when the magnetic
field and the friction are sufficiently strong. The suppression
owing to the friction 
occurs even along the field line. If the magnetic field and the
friction are not so strong, the instability is not stabilized.
(2) The effect of the friction and the magnetic field is mainly
reduction of the growth rate of the thermal instability of weakly ionized
plasma.
(3) The effect of friction does not affect the critical wavelength
$\lambda _{\rm F}$ for the thermal instability. This yields that 
$\lambda _{\rm F}$ of the weakly ionized plasma
is not enlarged even when the magnetic field exists. We insist that
the thermal instability of the weakly ionized plasma in the magnetic
field can grow up even at the small length scale where
the instability under the
assumption of the one fluid plasma can not grow owing to the
stabilization by the magnetic field.
(4) The wavelength of the maximum growth rate of the instability
shifts shortward according to the decrement of the growth rate. This is
because the friction is effective at rather larger scale. Therefore,
smaller structures are expected to appear 
than those without the ion-neutral friction.
Our results indicate the friction with the magnetic field affects the
morphology and evolution of the interstellar matter. In summary, the
ion-neutral friction is important for the evaluation of the thermal
instability in weakly ionized plasma along and perpendicular to the
magnetic field.

\end{abstract}




\keywords{
instabilities --- ISM: evolution}

\section{INTRODUCTION}

The recent progress of the interstellar medium (ISM) observations
has established that the small and tiny scale structures of ISM is 
very ubiquitous. Historically, the initial observational target is
the neutral phase of ISM.
Indeed, the tiny scale structures are
discovered by 21 cm absorption line observations against quasars with
VLBI techniques \citep{Di76}. The results are confirmed by
\citet{Dia89}. To look for the small scale structures in the cold
neutral phase of the ISM furthermore, 
the observations of 21 cm absorption lines
have been also performed against pulsars \citep{Fr94}. \citet{Me96}
observe optical spectral lines to find 
small structures against close binary stars.
The images of the small-scale H \textsc{i} have been
taken with the MERLIN array \citep{Da96}. These results are established
by higher angular resolution observations with VLBA and VLA toward
quasars \citep{Fa98,FG01}, which show small clumps on the order of a few
AU in neutral Galactic H \textsc{i} clouds. Cold H \textsc{i} clouds
have significant structure 
in subparsec scales \citep{Gi00,Br05}. 
Such small and tiny scale structures has been detected
in the local interstellar medium as H \textsc{i} absorption lines, 
although their column density is very small \citep{BK05}. 

In addition to the small and tiny scale structures of H {\rm \textsc{i}} cloud,
the picture of the
planetary nebula NGC 7293 shows fine small structures and knots
\citep{Ro02,OD04}.
Even in the starforming regions, there are variety small scale structures.
For example, \citet{La95} observe some clumps with size
from 0.007 to 0.021 pc in the Taurus Molecular Cloud 1 (TMC1), in
particular, core D. The mass of these fragments is estimated to be
$\lesssim 0.01 - 0.15$ $M_{\odot}$. 
SPITZER has begun producing higher spatial
resolution mid-infrared maps \citep{Ch04},
and revealed the fine structure of the starforming region. 
We think that it is possible
for proto-brown dwarfs of $< 0.08$ $M_{\odot}$ to exist there.
A high resolution
observation in future is expected to reveal hidden, small structures, as
well as substellar objects with very small mass.
In any ways, the
tiny-scale structure seems to be ubiquitous, not associated with large
extinction \citep{He97}. 

To study the origin of 
the small-scale structures of ISM is important to understand
the evolution of and structure formation in ISM.
Especially, in the starforming regions,
the small, low-mass structures can relate to the
low-mass cutoff of the initial mass function,
the coagulation unit to form the massive stars,
and so on.
Then, we should investigate 
the physical origin of these small and tiny
structures in the partially ionized medium,
since the cold H {\rm \textsc{i}} and molecular clouds are weakly ionized.
By the way, it is not easy for the small and tiny scale structure
to form as a result of gravitational instability since
their size is much smaller than the Jeans length.
According to \citet{La95}, indeed, the small-scale structures
appear to be gravitationally unbound. This suggests that some
fragmentation mechanisms but pure Jeans gravitational instability may be
important in the clouds. We expect the thermal instability as
this mechanism, and revisit it in this paper. 

The basics of the thermal instability has been summarized
in \citet{F65}. When the
following condition is satisfied, the system is thermally unstable:
If the cooling becomes efficient as
the temperature decreases, the thermal energy gets lost more and more.
If the cooling becomes efficient during the fluid contraction, the
system shrinks on and on. 
Importantly, the critical length scale of the thermal instability
is smaller than that of the dynamical instability like 
the Jeans instability. That is,
even if a
system is stable against the gravitational instability, 
the system can become thermally unstable.
Then, the thermal instability
can be the physical origin of the 
smaller-scale formation than the dynamical instability. 
Indeed, \citet{KI02}
propose that the clumpiness in clouds emerges naturally from their
formation through the thermal instability. 
Their two-dimensional
calculations follow the fragmentation into small cloudlets that result
from the thermal instability in a shock-compressed layer. \citet{BL00}
investigate the cooling and fragmentation of optically thin gas with a
power-law cooling function. According to them, small-scale
perturbations have the potential to reach higher amplitude than
large-scale fluctuations. Thermal instability can be important
for the structure formation.

We sketch ISM
structure formation such as molecular clouds as follows. The hot
ionized ISM cools to be cold and weakly ionized clouds. Ionization
degree of ISM decreases as it evolves, and then the ISM fluid is often
partially ionized. In partially ionized fluid, an ion component and
neutral component interact each other, exchanging the momentum (i.e.,
ion-neutral friction). Especially, the partially ionized plasma in a
magnetic field could not be treated as one fluid.
This is because although the ion
component in the partially ionized plasma is directly influenced by a
magnetic field, the neutral component does not directly feel a
magnetic field. In weakly ionized fluid, the neutral component is
affected by a magnetic field via the ion-neutral friction. For example,
ambipolar diffusion takes an important role in dynamical evolution
owing to the gravitational instability of ISM into protostar \citep{MeSp56,
Na76}. If the fluid frozen in a magnetic field contracts without
ambipolar diffusion, the magnetic field becomes too large and suppresses
the growth by the magnetic pressure and tension. In addition,
\citet{Na79} points out the possibility that ISM with sufficient
magnetic flux quasistatically evolves into protostar. 
He also notices that the
time scale of plasma drift depends on the ionization degree.
The relation
among the ion-neutral friction, the magnetic field, and the amount of
the ion component is important to understand the detailed process. 


We here emphasize that the MHD approximation is not always applicable 
for our purpose in this paper. The thermal instability is effective in small
structure formation, during which a neutral component is not always
enough frozen in an ion component dynamically to be treated as one
fluid. 
There is a possibility that the ion component can drift away
owing to ambipolar diffusion, and that the system can contract owing to
thermal instability. Then, 
in the present paper, 
to study how the ion-neutral drag
and magnetic field influence the growth of the thermal instability, we 
treat the plasma as the two fluid of ions and neutrals. We assume that the
ionization degree is very small, because we are interested in the final
stage of the formation of a molecular cloud, in which the ionization
degree decreases very much. We focus on the condensation mode of
thermal instability, since it is important in structure formation, rather
than the oscillation (overstable) mode. In \S2, the problem is formulated.
The property of the dispersion relation of the instability is presented
in \S3. Several discussions relating to the instability are found in \S4.
Applicability of our results to cold H {\rm \textsc{i}} medium is briefly examined
in \S5, and then
we summarize the paper in \S6.

\section{BASIC EQUATIONS AND DISPERSION RELATION}
\label{S_BasicEq}

\subsection{Basic Equations}
The basic equations for ion-neutral two component fluid are as follows. 
The continuity equation, equation of motion, energy equation, 
and equation of state are
\begin{eqnarray}
\frac{\partial \rho_{\rm n}}{\partial t} + \nabla (\rho_{\rm n} \mbox{\boldmath$v$}_{\rm n}) &=& 0
\label{cont_n}
,\\
\rho_{\rm n} \left[\frac{\partial \mbox{\boldmath$v$}_{\rm n}}{\partial t} +  (\mbox{\boldmath$v$}_{\rm n} \cdot \nabla ) \mbox{\boldmath$v$}_{\rm n} \right] &=& - \nabla p_{\rm n} - \rho_{\rm n} \nu_{\rm ni} (\mbox{\boldmath$v$}_{\rm n} - \mbox{\boldmath$v$}_{\rm i})
,
\label{eqmotion_n}
\\
\frac{1}{\gamma-1} \left[ \frac{\partial p_{\rm n}}{\partial t} + (\mbox{\boldmath$v$}_{\rm n} \cdot \nabla) p_{\rm n} \right] - \frac{\gamma}{\gamma - 1} \frac{p_{\rm n}}{\rho_{\rm n}} \left[ \frac{\partial \rho_{\rm n}}{\partial t} + (\mbox{\boldmath$v$}_{\rm n} \cdot \nabla) \rho_{\rm n} \right] &=& - \rho_{\rm n} \Lambda_{\rm n} + \nabla (K_{\rm n} \nabla T_{\rm n}) 
,
\label{eq:energy_n}
\end{eqnarray}
and
\begin{eqnarray}
p_{\rm n} &=& \frac{\cal R}{\mu_{\rm n}} \rho_{\rm n} T_{\rm n}
,
\end{eqnarray}
for a neutral component. 
Those for an ion component are
\begin{eqnarray}
\frac{\partial \rho_{\rm i}}{\partial t} + \nabla (\rho_{\rm i} \mbox{\boldmath$v$}_{\rm i}) &=& 0
,\\
\rho_{\rm i} \left[\frac{\partial \mbox{\boldmath$v$}_{\rm i}}{\partial t} + (\mbox{\boldmath$v$}_{\rm i} \cdot \nabla ) \mbox{\boldmath$v$}_{\rm i} \right] &=& - \nabla p_{\rm i} - \rho_{\rm n} \nu_{\rm ni} (\mbox{\boldmath$v$}_{\rm i} - \mbox{\boldmath$v$}_{\rm n}) 
\nonumber \\
& & {}
+ \frac{1}{4 \pi} (\nabla \times \mbox{\boldmath$B$}) \times \mbox{\boldmath$B$}
,
\label{eqmotion_I}
\\
\frac{1}{\gamma-1} \left[ \frac{\partial p_{\rm i}}{\partial t} + (\mbox{\boldmath$v$}_{\rm i} \cdot \nabla) p_{\rm i} \right] - \frac{\gamma}{\gamma - 1} \frac{p_{\rm i}}{\rho_{\rm i}} \left[ \frac{\partial \rho_{\rm i}}{\partial t} + (\mbox{\boldmath$v$}_{\rm i} \cdot \nabla) \rho_{\rm i} \right] &=& - \rho_{\rm i} \Lambda _{\rm i} + \nabla (K_{\rm i} \nabla T_{\rm i})
,
\end{eqnarray}
\begin{eqnarray}
p_{\rm i} &=& \frac{\cal R}{\mu_{\rm i}} \rho_{\rm i} T_{\rm i}
,
\end{eqnarray}
and the induction equation is 
\begin{eqnarray}
\frac{\partial \mbox{\boldmath$B$}}{\partial t} &=& \nabla \times (\mbox{\boldmath$v$}_{\rm i} \times \mbox{\boldmath$B$})
\label{inductioneq}
.
\end{eqnarray}
Here, the subscript, n, denotes neutral, and subscript, i, does ion. 
Variables $\rho$, $\mbox{\boldmath$v$}$, $p$, $T$,
 and $\mbox{\boldmath$B$}$ express density, velocity, pressure, temperature, and a magnetic field, respectively. 
The cooling function $\Lambda$ is defined as energy losses minus energy gains per gram per second. 
In addition, the coefficient of thermal conductivity is expressed by $K$,
 the gas constant by ${\cal R}$, and the mean molecular weight by $\mu$.

The second terms on the right-hand sides of equations (\ref{eqmotion_n})
 and (\ref{eqmotion_I}) mean the effect of dragging through the neutral-ion collision, similarly to \citet{Wa04}.
The neutral-ion collision frequency $\nu_{\rm ni}$ is expressed as $ \gamma _\nu \rho_{\rm i}$, where $\gamma _\nu$ is the collision rate coefficient per unit mass, 
\begin{equation}
\gamma _\nu \equiv \frac{\langle \sigma v \rangle}{m_{\rm i} + m_{\rm n}}.
\end{equation}
Here, $\sigma$ is the collision cross-section,
 $v$ is the relative velocity between ion and neutral components,
 $m$ is mass, and the angular bracket of $\langle \rangle$ expresses a mean value over all velocities.
It is pointed out 
that the larger the friction is, 
the better the one fluid approximation of the partially ionized plasma is 
like a classical MHD approximation.


The effect of the heat flow between the two components due to the difference of the temperatures 
is neglected as second order terms because we assume the temperatures of two components under the equilibrium are identical. 
In addition, the frictional heating is also neglected as second order terms 
because the mutual velocity between the two components is zero in the equilibrium.

\subsection{Dispersion Relation}
We solve the perturbed basic equations, considering the first order. 
The unperturbed state is supposed to be infinite, uniform, static, and isothermal in equilibrium.
It is described with subscript 0, 
such as $\rho_{\rm n} = \rho_{{\rm n}0}$, 
$p_{\rm n} = p_{{\rm n}0}$,
 $\mbox{\boldmath$v$}_{\rm n} = \mbox{\boldmath$v$}_{{\rm n} 0} = 0$,
 $T_{\rm n} = T_{{\rm n}0}$, 
$\rho_{\rm i} = \rho_{{\rm i}0}$, 
$p_{\rm i} = p_{{\rm i}0}$,
 $\mbox{\boldmath$v$}_{\rm i} = \mbox{\boldmath$v$}_{{\rm i} 0} = 0$,
 $T_{\rm i} = T_{{\rm i}0}$, and 
$\mbox{\boldmath$B$} = \mbox{\boldmath$B$}_{0}$.
The magnetic field $\mbox{\boldmath$B$}_{0}$ in equilibrium can be 
taken as $\mbox{\boldmath$B$}_{0} = (0, 0, B_{0})$ for simplicity.
The net transfer of thermal energy for each component in equilibrium is zero, such as 
$ \Lambda _{\rm n} (\rho _{{\rm n} 0}, T_{{\rm n} 0}) =0$  
and 
$ \Lambda _{\rm i} (\rho _{{\rm i} 0}, T_{{\rm i} 0}) =0$.

Someone might think our unperturbed state is inadequate since the size of the unperturbed medium is infinite.
However, in our so-called WKB framework, we can state how our unperturbed condition is useful. 
As long as the WKB approximation is adopted, the whole of the system should be large.
This can correspond to the fact that the scale length of the galactic gas disk is about 100 pc, which is always much 
longer than the so-called Field length, $\lambda _{\rm F}$. 
That is, since the interesting wave-length of the instability for the structure formation is smaller than the thickness 
of the galactic disk, our study is meaningful.

We take the perturbation as 
\begin{equation}
a(\mbox{\boldmath$r$}, t) = a_{1} 
\exp (nt + i \mbox{\boldmath$k$} \cdot \mbox{\boldmath$r$}),
\label{perturb}
\end{equation}
according to \citet{F65}, 
where $a_{1}$ is the amplitude of perturbation, 
$n$ is the growth rate of perturbation,
and $\mbox{\boldmath$k$}$ is the wave number of perturbation.
When $n$ is real and positive, the perturbation grows exponentially
as a condensation mode. If $n$ is real and negative,
the perturbation damps. 
When $n$ is complex number, the system is oscillatory growing or damping.
Of course, when $\Re[n]$ is positive, the system is so-called overstable.

In order to know how the magnetic field affects on weakly ionized fluid, 
we mainly investigate the condensation mode, 
because we are interested in the formation and evolution of molecular clouds. 
We assume that 
\begin{equation}
\mbox{\boldmath$B$}_{0}
\perp
\mbox{\boldmath$k$}
~~
{\rm and}
~~
\mbox{\boldmath$k$} \parallel
\mbox{\boldmath$v$}_{{\rm i} 1}
\parallel
\mbox{\boldmath$v$}_{{\rm n} 1}
\label{itijigen}
\end{equation}
for comfort. 
Due to the induction equation (\ref{inductioneq}),
these assumptions lead to
\begin{equation}
\mbox{\boldmath$B$}_{0} \parallel
\mbox{\boldmath$B$}_{1}
,
\label{itijigen02}
\end{equation}
[see equation (\ref{linear_eq_induc}) in Appendix \ref{Ap_del}]. 
It is emphasized that the assumed magnetic field is perpendicular to the fluid motion. 


Here, it is noticed that we can examine some other modes, 
which do not satisfy the condition (\ref{itijigen}) and (\ref{itijigen02}),
when $\mbox{\boldmath$B$}_{0} = 0$. 
This means $\mbox{\boldmath$B$}_{1} = 0$, 
owing to the induction equation (\ref{inductioneq}).
In other words, we can study the property of the dispersion
relation when the magnetic field does not work effectively.
This is interesting point in our analysis since
those modes with $\mbox{\boldmath$B$}_{0} = 0$ 
yield understanding of the pure effect by the ion-neutral friction. 
It is also noted that there is no thermal instability 
in a pure Alfv\'{e}n mode
which satisfies $\mbox{\boldmath$v$} \perp \mbox{\boldmath$k$}$, 
because the Alfv\'{e}n mode is effectively incompressible,
 and there is no net thermal exchange,
 even if the fluid is compressible.

\bigskip

In this paper, since we are interested in the structure formation
from H {\rm \textsc{i}} medium, 
the fluid is assumed to be weakly ionized, i.e., 
$\rho _{{\rm i}0} \ll \rho _{{\rm n}0}$.
We also assume that temperatures of neutral 
and ion components are the same in the equilibrium, 
i.e., $T_{{\rm n}0} = T_{{\rm i}0}$,
because they are well mixed and collide enough to 
have the same temperature for fewness of ion component. 
Using these conditions and replacing 
$ \vert \mbox{\boldmath$k$} \vert = k$,
here is the dispersion relation 
(see Appendix \ref{Ap_del} for the derivation in detail): 
\clearpage
\begin{eqnarray}
\lefteqn{
\left[
n^3 
+ n^2 \left(v_{\rm sn} k_{\rm nT} + v_{\rm sn} k \frac{k}{k_{\rm n K}} \right)
+ n (v_{\rm sn} k)^2 
+ \frac{1}{\gamma} (v_{\rm sn} k)^2
\left( v_{\rm sn} k_{\rm nT} - v_{\rm sn} k_{\rm n \rho}
+ v_{\rm sn} k \frac{k}{k_{\rm n K}} \right)  
\right] 
}
\nonumber \\
& & {}
\lefteqn{
\times 
\left[
n^3 
+ n^2 \left(v_{\rm si} k_{\rm iT} + v_{\rm si} k \frac{k}{k_{\rm i K}} \right)
+ n (v_{\rm si} k)^2 
+ n (v_{\rm A} k)^2
\right.
}
\nonumber \\
& & {} 
\lefteqn{\left.
\hspace{1.1cm}
+ \frac{1}{\gamma}
\left( v_{\rm si} k_{\rm iT} + v_{\rm si} k \frac{k}{k_{\rm i K}} \right)
\left( v_{\rm si}^2 k^2 + \gamma v_{\rm A}^2 k^2 \right)
- \frac{1}{\gamma} (v_{\rm si} k)^2 (v_{\rm si} k_{\rm i \rho})
\right]
}
\nonumber \\
& = &
-\nu _{{\rm ni} 0} \frac{1}{\chi}
 \cdot \left\{
n^5
+ n^4 
\left( 
v_{\rm si} k_{\rm iT} + v_{\rm si} k \frac{k}{k_{\rm i K}} 
+ v_{\rm sn} k_{\rm nT} + v_{\rm sn} k \frac{k}{k_{\rm n K}} 
\right)
\right.
\nonumber \\
& & {}
+ n^3 
\left[ 
v_{\rm sn} k \frac{k}{k_{\rm nK}} 
v_{\rm si} k \frac{k}{k_{\rm iK}}
+ v_{\rm sn} k \frac{k}{k_{\rm nK}} v_{\rm si} k_{\rm iT}
+ v_{\rm si} k \frac{k}{k_{\rm iK}} v_{\rm sn} k_{\rm nT}
+ v_{\rm sn} k_{\rm nT} v_{\rm si} k_{\rm iT}
\right.
\nonumber \\
& & {}
\left.
\hspace{1.5cm}
+ (v_{\rm sn} k)^2 + (v_{\rm A} k)^2 \chi
\right]
\nonumber \\
& & {}
+ n^2 
\left[
\frac{1}{\gamma}
(v_{\rm sn} k)^2 
(v_{\rm sn} k_{\rm nT} + v_{\rm sn} k \frac{k}{k_{\rm n K}})
- \frac{1}{\gamma} (v_{\rm sn} k)^2 v_{\rm sn} k_{\rm n \rho}
+ (v_{\rm sn} k)^2 
(v_{\rm si} k_{\rm iT} + v_{\rm si} k \frac{k}{k_{\rm i K}})
\right.
\nonumber \\
& & {}
\hspace{1.5cm}
\left.
- \frac{1}{\gamma} \chi (v_{\rm si} k)^2 v_{\rm si} k_{\rm i \rho}
+ (v_{\rm A} k)^2 \chi 
\left( 
v_{\rm si} k_{\rm iT} + v_{\rm si} k \frac{k}{k_{\rm i K}} 
+ v_{\rm sn} k_{\rm nT} + v_{\rm sn} k \frac{k}{k_{\rm n K}} 
\right)
\right]
\nonumber \\
& & {} 
+ n 
\left[
( v_{\rm A}^2 k^2 \chi + \frac{1}{\gamma} v_{\rm sn}^2 k^2 )
\right.
\nonumber \\
& & {}
\hspace{1.5cm}
\left(
v_{\rm sn} k \frac{k}{k_{\rm n K}} v_{\rm si} k \frac{k}{k_{\rm i K}}
+ v_{\rm sn} k \frac{k}{k_{\rm n K}} v_{\rm si} k_{\rm iT} 
+ v_{\rm si} k \frac{k}{k_{\rm i K}} v_{\rm sn} k_{\rm nT}
+ v_{\rm sn} k_{\rm nT} v_{\rm si} k_{\rm iT}
\right)
\nonumber \\
& & {}
\left.
\left.
- \frac{1}{\gamma} \chi (v_{\rm si} k)^2 v_{\rm si} k_{\rm i \rho}
(v_{\rm sn} k_{\rm nT} + v_{\rm sn} k \frac{k}{k_{\rm n K}})
- \frac{1}{\gamma} (v_{\rm sn} k)^2 v_{\rm sn} k_{\rm n \rho}
( v_{\rm si} k_{\rm iT} + v_{\rm si} k \frac{k}{k_{\rm i K}} )
\right]
\right\}
,
\label{dis-rela}
\end{eqnarray}
where
\begin{equation}
v_{\rm sn}^2 = \gamma \frac{p_{{\rm n}0}}{\rho_{{\rm n}0} },
 ~~
v_{\rm si}^2 = \gamma \frac{p_{{\rm i}0}} { \rho_{{\rm i}0} },
 ~~ 
v_{\rm A}^2 = \frac{B _0 ^2}{4 \pi \rho _{{\rm i}0} },
\end{equation}

\begin{equation}
k_{\rm n \rho}=
\frac{\mu_{\rm n}(\gamma -1) \rho_{{\rm n}0} \Lambda _{\rm n \rho} }
{{\cal R} v_{\rm s n} T_{{\rm n} 0}}, ~~
k_{\rm n T}=
\frac{\mu_{\rm n}(\gamma -1) \Lambda _{\rm n T} }
{{\cal R} v_{\rm s n}}, ~~
k_{\rm n K}=
\frac{{\cal R} v_{\rm s n} \rho _{{\rm n} 0}}
{\mu_{\rm n}(\gamma -1) K_{{\rm n}0}},
\label{eq:det_para_normalization_three_k_for_neutral}
\end{equation}

\begin{equation}
k_{\rm i \rho}=
\frac{\mu_{\rm i}(\gamma -1) \rho_{{\rm i}0} \Lambda _{\rm i \rho} }
{{\cal R} v_{\rm s i} T_{{\rm i} 0}}, ~~
k_{\rm i T}=
\frac{\mu_{\rm i}(\gamma -1) \Lambda _{\rm i T} }
{{\cal R} v_{\rm s i}}, ~~
k_{\rm i K}=
\frac{{\cal R} v_{\rm s i} \rho _{{\rm i} 0}}
{\mu_{\rm i}(\gamma -1) K_{{\rm i}0}},
\end{equation}
and
\begin{equation}
\nu_{\rm ni 0} \equiv \gamma _\nu \rho_{\rm i 0}, ~~
\chi \equiv \frac{\rho _{{\rm i}0}} {\rho _{{\rm n} 0} }
.
\end{equation}
In addition, 
\begin{equation}
\Lambda _{\rm n \rho} \equiv 
\left( \frac{\partial \Lambda _{\rm n}}{\partial \rho _{\rm n}}
\right) _{T _{\rm n}} , ~~
\Lambda _{\rm n T} \equiv 
\left( \frac{\partial \Lambda _{\rm n}}{\partial T_{\rm n}}
\right) _{\rho _{\rm n}} , ~~
\Lambda _{\rm i \rho} \equiv 
\left( \frac{\partial \Lambda _{\rm i}}{\partial \rho _{\rm i}}
\right) _{T _{\rm i}} , ~~
\Lambda _{\rm i T} \equiv 
\left( \frac{\partial \Lambda _{\rm i}}{\partial T_{\rm i}}
\right) _{\rho _{\rm i}} ,
\end{equation}
and they are evaluated in the equilibrium state.

The dispersion relation (\ref{dis-rela}) is 
 the equation of sixth order of $n$.
This dispersion relation connotes the two sets of three modes.
One of the three modes in each set is condensation mode, 
and the other two modes are oscillation modes
originally, if there is no friction effect.
The first bracket on the left-hand side in equation (\ref{dis-rela})
is the dispersion relation of three modes of only neutral fluid 
(\ref {crite_cond}).
The second in equation (\ref{dis-rela}) 
is the dispersion relation of only ionized fluid in the magnetic field 
(\ref{crite_ion}).
The right-hand side of equation (\ref{dis-rela}) expresses 
the momentum exchange through the ion-neutral drag.
The properties of the thermal instability of one fluid approximation is also
briefly mentioned in Appendix \ref{subsec:nature_dispersion_relation}.

\section{RESULTS}
\label{S_Results}

\subsection{Critical Wavelength of the Condensation Mode}
\label{subsec:results:critical_wavelength}

Observing the dispersion relation (\ref{dis-rela}), we obtain the first
result that a critical wavelength for the instability in a condensation
mode is not affected by the friction. The critical wavelength is
evaluated if we let $n$ be zero, because the growth rate of $n$ is zero
at the critical wavelength. Thus, letting $n =0$, we find the
right-hand side of equation (\ref{dis-rela}), which denotes the effect
of the ion-neutral friction, diminishes. That is, the effect of the
friction does not play any roles to determine the critical wavelength
of the condensation mode. The criterion of the instability for the
partially ionized plasma is estimated as if there is no friction.
Although the critical wavelength is not affected, the friction affects
the growth rate of the thermal instability. It is noted that the
criterion of the overstability modes is not derived by the same way like
the condensation mode since $n$ is imaginary.

\subsection{Growth Rate of Condensation Mode of the Instability}

\subsubsection{Description of the Conditions}

The dispersion relation (\ref{dis-rela}) is numerically solved
by means of MATHEMATICA, 
after the normalization (see Appendix \ref{Ap_nor}).
We study the condensation mode which is important on structure formation.
To present the numerical dispersion relations,
we take the following parameters : 
$ \gamma = 5/3$, 
$\alpha _{\rm n} \equiv k_{\rm nT} / k_{\rm n \rho} = 1/2$, 
$\alpha _{\rm i} \equiv k_{\rm iT} / k_{\rm i \rho} = 1/2$, 
$\beta _{\rm n} \equiv k_{\rm n \rho} / k_{\rm n K} = 0.01$, 
$\beta _{\rm i} \equiv k_{\rm i \rho} / k_{\rm i K} = 0.01$, 
for the comparison to \citet{F65}.
We set these parameters to be fixed 
because we are concerned for the basic nature
of two-component fluid and compare the results with one-component fluid.
We assume that
$ \mu _{\rm n} = 1$, $\mu _{\rm i} = 1/2$, 
for e.g., only H atom plasma, 
$ \chi = 0.01 $, and
$ k_{\rm i \rho} = k_{\rm n \rho}$. 
Under these conditions, 
the dispersion relation (\ref{dis-rela}) is evaluated
for the various quantities of 
$v _{\rm A} / v _{\rm sn}$ which corresponds 
to the strength of the magnetic field, and 
$\nu _{{\rm ni} 0} / ( k _{{\rm n} \rho} v _{\rm sn})$
 which corresponds to the strength of the friction.

Before showing the precise results of the numerical dispersion
relations, we examine the physical meaning of the axis of the following figures.
In each figure, 
the horizontal axis is the wave number $k$ in units of $ k_{\rm n \rho} $. 
The vertical axis corresponds to the growth rate $\Re [n]$ 
in units of $ k_{\rm n \rho} v_{\rm sn} $.
These normalizations are adopted to denote the characteristic 
property of the thermal instability. 
The wavelength is taken as $\lambda$.
We shall start considering the physical meaning of $k_{\rm n \rho}$.
According to equation 
(\ref{eq:det_para_normalization_three_k_for_neutral}) 
with approximation of 
$\rho_{{\rm n}0} \Lambda _{\rm n \rho} \sim \Lambda _{\rm n}$, 
\begin{eqnarray}
k_{\rm n \rho}
& = & 
\frac{\mu_{\rm n}(\gamma -1) \rho_{{\rm n}0} \Lambda _{\rm n \rho} }
{{\cal R} v_{\rm s n} T_{{\rm n} 0}}
\\ \nonumber
& \sim & {}
\frac{\rho_{{\rm n}0} \Lambda _{\rm n}}{p_{{\rm n}0}} \cdot \frac{1}{v_{\rm s n}}
,
\end{eqnarray}
where $p_{{\rm n}0} / \rho_{{\rm n}0} \Lambda _{\rm n}$ expresses the
cooling time. Then, we find $k_{\rm n \rho} ^{-1}$ is interpreted as the
distance for the fluid to acoustically move within the cooling time. In
other words, it is interpreted as the length scale in which 
the fluid can respond
to the perturbation. Therefore, regarding the horizontal axis, 
$(k /k_{\rm n \rho} )^{-1} \sim \lambda k_{\rm n \rho}$ is interpreted as the
scalelength in units where the fluid can respond to the thermal
instability within the cooling time (i.e., the growing time). 
We get some physical intuition once the above normalization is adopted. 
For example, we get physical intuition when 
$(k / k_{\rm n \rho})^{-1} \gg 1$, 
the fluid is hard to respond to the thermal
instability. This is because the system is too large and the
acoustic wave which carries the disturbance can not travel through the
whole system within the cooling time. 
Then, the growth rate at
$(k/k_{\rm n \rho})^{-1} \gg 1$ is reduced down in each figure. 
When 
$(k/k_{\rm n \rho})^{-1} < 1$, the fluid is easy to respond to the
thermal instability.

Similarly, regarding the vertical axis, 
$\Re [n] /k_{\rm n \rho} v_{\rm sn} $ is 
interpreted as $n$ multiplied by the
cooling time. In other word, it is interpreted as the cooling time
divided by the time scale for the growth of the instability. Then, the
vertical axis means how fast the instability can grow up compared
with the cooling time. The thermal instability is due to the cooling,
then the value of $\Re [n] / k_{\rm n \rho} v_{\rm sn} $ should be less
than unity as presented in each figure.

\subsubsection
{Various Conditions of the Friction and Magnetic Field}
\label{subsubsec:results:various_conditions}

\begin{figure}[htbp]
	\plotone{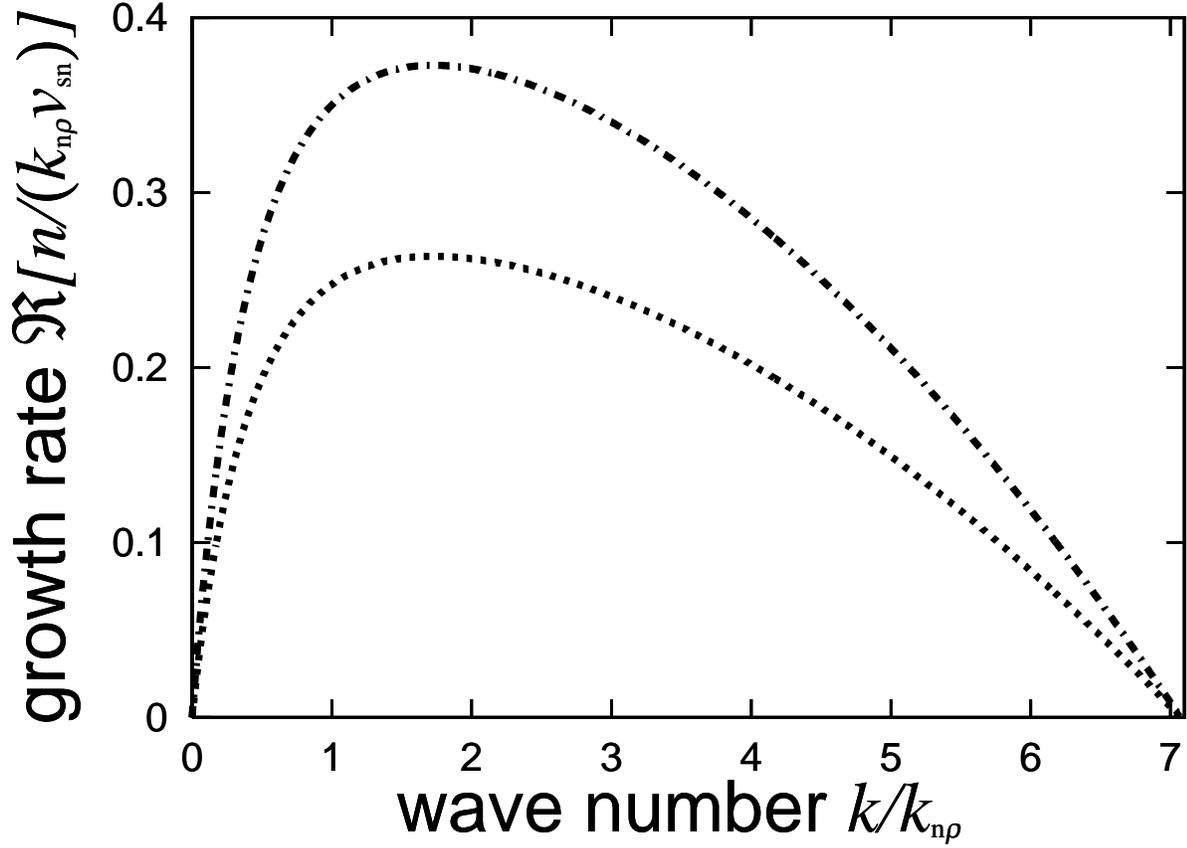}
	\caption
	{
	Growth rates of each condensation mode of the ion component (the dot-dashed curve) 
	and the neutral component (the dashed one), 
	when $ B=0$ and $\nu _{{\rm ni}0} = 0$. 
	The dot-dashed curve corresponds to fully ionized fluid
 and dashed one to totally neutral fluid. 
	The horizontal axis is the normalized wave number $ k / k_{\rm n \rho} $.
    The vertical axis corresponds to the normalized growth rate 
    $ \Re [n / ( k_{\rm n \rho} v_{\rm sn} ) ]$.
    }
	\label{B0C0}
\end{figure}

Figure \ref{B0C0} displays the dispersion relation of the real part of
$n$ when there exists neither the magnetic field nor friction. A
dot-dashed curve corresponds to a mode characterized by $\mu_{\rm i}$
and a dashed one to that of $\mu_{\rm n}$.
No friction means that the motion of the ion and
neutral components are independent completely, and thus a dot-dashed
curve corresponds to a single ionized fluid and a dashed one to a simple
neutral fluid. The difference between the two curves comes from the
mean molecular weight. This result follows \citet{F65}.

\begin{figure}[htbp]
	\plotone{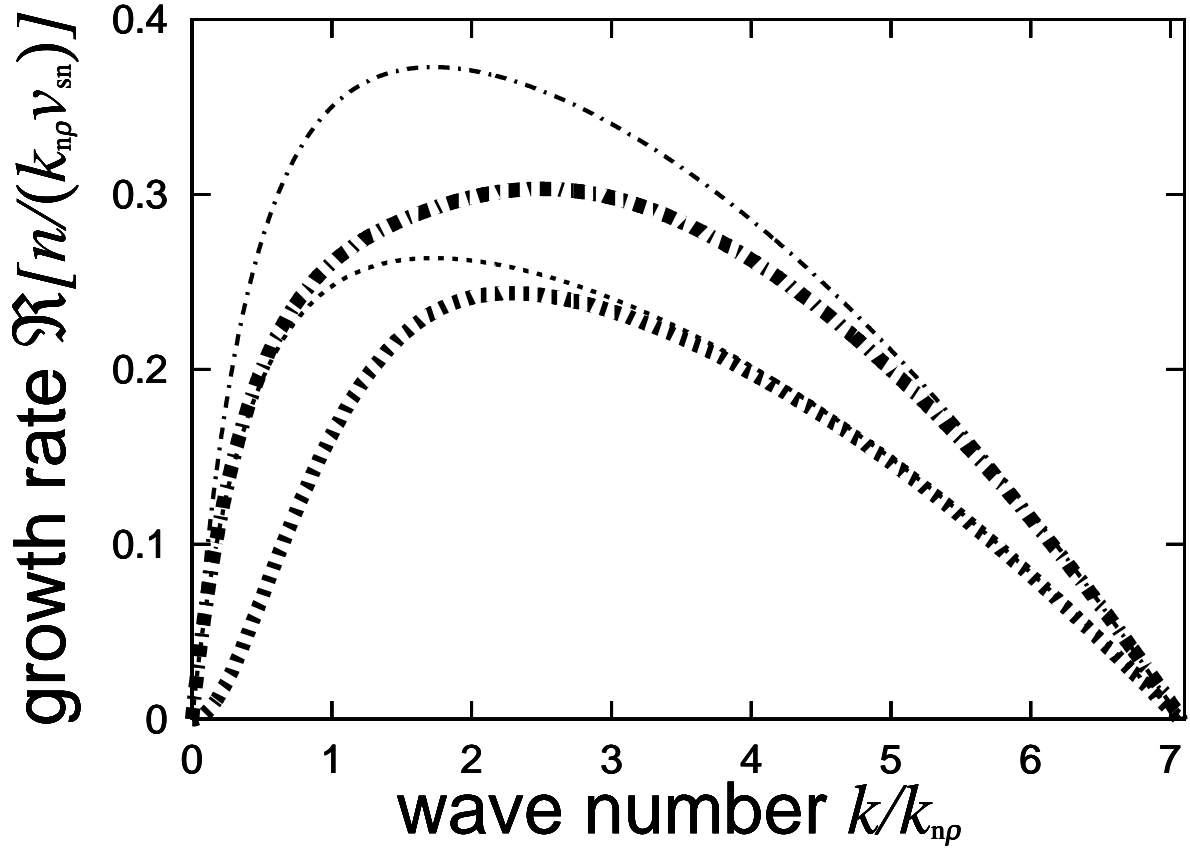}
	\caption
	{
	Growth rates of the condensation mode for the instability 
	characterized by $\mu_{\rm n}$  (the thick dashed curve)
	  and $\mu_{\rm i}$ (the thick dot-dashed curve), 
	when $ v_{\rm A} / v_{\rm sn} = 0$ and $\nu _{{\rm ni}0} / ( k_{\rm n \rho} v_{\rm sn} ) = 0.03$. 
	The horizontal axis is the normalized wave number $ k / k_{\rm n \rho} $.
    The vertical axis corresponds to the normalized growth rate 
    $ \Re [n / ( k_{\rm n \rho} v_{\rm sn} ) ]$.
	This figure also corresponds to the case
	that the fluid moves along the magnetic field line.
    For comparison, there are also displayed
     condensation modes of the ion component (the thin dot-dashed curve) 
	and  the neutral component (the thin dashed curve)
	when $ B=0$ and $\nu _{{\rm ni}0} = 0$, as well as Figure \ref{B0C0}.
    }
	\label{B0C003}
\end{figure}

Figure \ref{B0C003} shows the growth rate when there is no magnetic
field but some friction [i.e., $ v_{\rm A} / v_{\rm sn} = 0$, $\nu
_{{\rm ni}0} / ( k_{\rm n \rho} v_{\rm sn} ) = 0.03$]. This condition
yields understanding of the pure effect of the friction. When there
exists the friction, a pure mode of an ion component or a neutral one
does not exist. Therefore, interpretation is required on what a curve
in the figure means when there exists the friction. 
The resultant modes with the friction are displays as two thick lines.
For
comparison, the two thin lines when there exists neither a magnetic field
nor friction in Figure \ref{B0C0} are also plotted.
A thick dot-dashed curve
originates from an ion component mode of Figure 1.
A thick dashed curve originates from
an neutral component of Figure 1.
These interpretations are reasonable since
the effect of mean molecular weight of each component 
is imprinted in each dispersion relation.
The difference between the Figure 1 and 2 is that the 
growth rate at the larger scale decreases, compared
with the original modes of Figure 1.
This is because the suppression by the friction is more efficient at rather
larger scale. The longer distance particles move, the more
frequent the collision from which the friction originates is. 
It is noticed that the current model corresponds to the case
that the fluid moves along the field line if there exists the
magnetic field. 
Figure \ref{B0C003} shows that
only the friction suppresses the instability along the field line, 
while the magnetic field does not.

\begin{figure}[htbp]
	\plotone{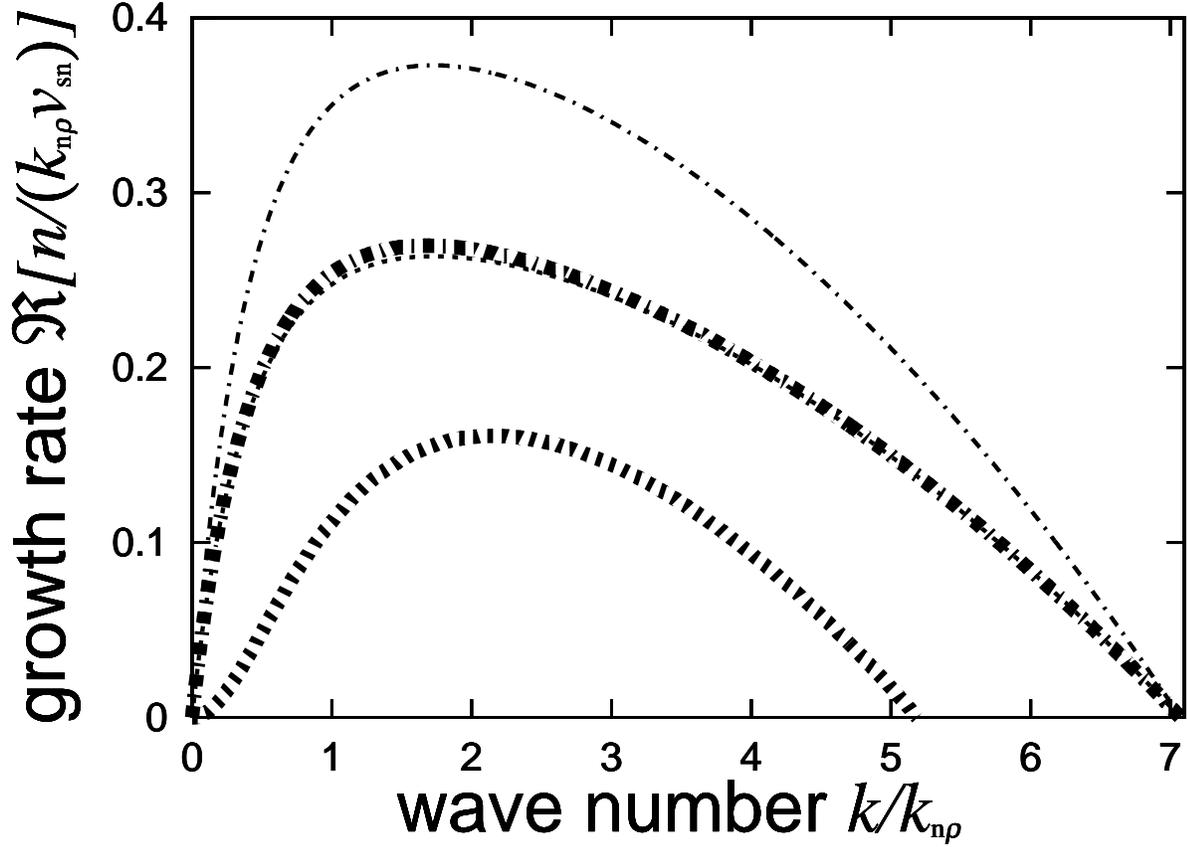}
	\caption
	{
	Growth rates of the condensation mode for the instability
	characterized by $\mu_{\rm n}$ (the thick dot-dashed curve)
	  and $\mu_{\rm i}$ (the thick dashed curve), 
	when $ v_{\rm A} / v_{\rm sn} = 0.6$ and $\nu _{{\rm ni}0} / ( k_{\rm n \rho} v_{\rm sn} ) = 0.03$. 
	The horizontal axis is the normalized wave number $ k / k_{\rm n \rho} $.
    The vertical axis corresponds to the normalized growth rate 
    $ \Re [n / ( k_{\rm n \rho} v_{\rm sn} ) ]$.
    The difference between Figures \ref{B0C003} and \ref{B06C003}
     is the magnetic field strength. 
    For comparison, there are also displayed
     condensation modes of the ion component (the thin dot-dashed curve) 
	and  the neutral component (the thin dashed curve)
	when $ B=0$ and $\nu _{{\rm ni}0} = 0$, like Figure 2.
    }
	\label{B06C003}
\end{figure}

Figure \ref{B06C003} shows the growth rate when the friction is the same
as Figure \ref{B0C003}, but there is the finite magnetic field [i.e., $
v_{\rm A} / v_{\rm sn} = 0.6$, $\nu _{{\rm ni}0} / ( k_{\rm n \rho}
v_{\rm sn} ) = 0.03$ ]. 
For comparison, the
two thin curves which correspond to the results of 
Figure \ref{B0C0} are also displayed, as well as Figure \ref{B0C003}.
A thick dashed curve is a mode, which is characterized by 
$\mu_{\rm i}$.
The reason why the curve is interpreted to relating to $\mu_{\rm i}$
is
that the critical wavelength is enlarged owing to the stabilization of
the magnetic field [see the criterion (\ref{crite_ion})]. 
That is, the magnetic field suppresses a kind of mode of
the thermal instability.
A thick
dot-dashed curve is characterized by $\mu_{\rm n}$.
The critical wavelength
of this mode is not affected by the magnetic field. 
Thus, the mode characterized by $\mu_{\rm n}$
is hardly affected by the field. 
This occurs since the amount of the neutral
component significantly dominates that of the ion component.

\begin{figure}[htbp]
	\plotone{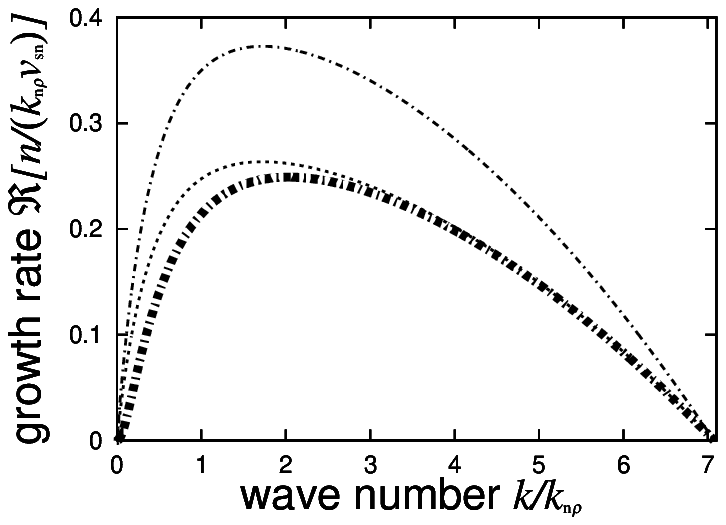}
	\caption
	{
	Growth rates of the condensation mode for the instability
	characterized by $\mu_{\rm n}$ (the thick dot-dashed curve), 
	when $ v_{\rm A} / v_{\rm sn} = 600$ 
	and $\nu _{{\rm ni}0} / ( k_{\rm n \rho} v_{\rm sn} ) = 0.3$. 
	The mode of $\mu_{\rm i}$ is completely stabilized 
	by the magnetic field and does not appear in this figure.
	The horizontal axis is the normalized wave number $ k / k_{\rm n \rho} $.
    The vertical axis corresponds to the normalized growth rate 
    $ \Re [n / ( k_{\rm n \rho} v_{\rm sn} ) ]$.
    For comparison, there are also displayed
     condensation modes of the ion component (the thin dot-dashed curve) 
	and the neutral component (the thin dashed curve) 
	when $ B=0$ and $\nu _{{\rm ni}0} = 0$ like Figure 2.
    }
	\label{B600C03}
\end{figure}

\begin{figure}[htbp]
	\plotone{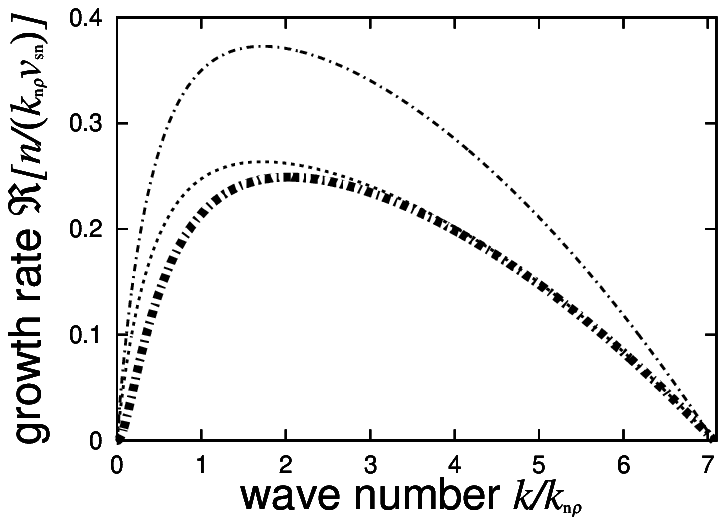}
	\caption
	{
	Growth rates of the condensation mode for the instability
	characterized by $\mu_{\rm n}$ (the thick dot-dashed curve), 
	when $ v_{\rm A} / v_{\rm sn} = 6000$ 
	and $\nu _{{\rm ni}0} / ( k_{\rm n \rho} v_{\rm sn} ) = 0.3$. 
	The mode of $\mu_{\rm i}$ is completely stabilized 
	by the magnetic field and does not appear in this figure.
	The horizontal axis is the normalized wave number $ k / k_{\rm n \rho} $.
    The vertical axis corresponds to the normalized growth rate 
    $ \Re [n / ( k_{\rm n \rho} v_{\rm sn} ) ]$.
    The difference between Figures \ref{B600C03} and \ref{B6000C03}
     is the magnetic field strength. 
    For comparison, there are also displayed
     condensation modes of the ion component (the thin dot-dashed curve) 
	and  the neutral component (the thin dashed curve) when $ B=0$ and $\nu _{{\rm ni}0} = 0$ like Figure 2.
    }
	\label{B6000C03}
\end{figure}

Figure \ref{B600C03} shows the growth rate
when the magnetic field is thousand times of that in Figure \ref{B06C003},
and friction is ten times of that in Figure \ref{B06C003}
[i.e., $ v_{\rm A} / v_{\rm sn} = 600$,
$\nu _{{\rm ni}0} / ( k_{\rm n \rho} v_{\rm sn} ) = 0.3$].
The large magnetic field completely stabilizes a mode
characterized by $\mu_{\rm i}$
so that this mode does not appear 
in this figure.
Thus, there is displayed only one mode of the thermal instability 
characterized by $\mu_{\rm n}$.
This is expressed by the thick dot-dashed curve.
For comparison,
the two thin curves are displayed as well as Figure \ref{B0C003}.
Figure \ref{B6000C03} shows the growth rate
when the magnetic field is ten times of that in Figure \ref{B600C03}
and friction is the same as in Figure \ref{B600C03}
[i.e., $ v_{\rm A} / v_{\rm sn} = 6000$,
$ \nu _{{\rm ni}0} / ( k_{\rm n \rho} v_{\rm sn} ) = 0.3$].
The large magnetic field completely stabilizes a mode
characterized by $\mu_{\rm i}$.
The thick dot-dashed curve expresses
the mode of thermal instability characterized by $\mu_{\rm n}$.
For comparison,
the two thin curves are displayed as well as Figure \ref{B0C003}. 
It is noted that Figures \ref{B600C03} and \ref{B6000C03} show that 
the suppression by the magnetic field is 
saturated (see subsection \ref{S_Discuss_Satu}).

\begin{figure}[htbp]
	\plotone{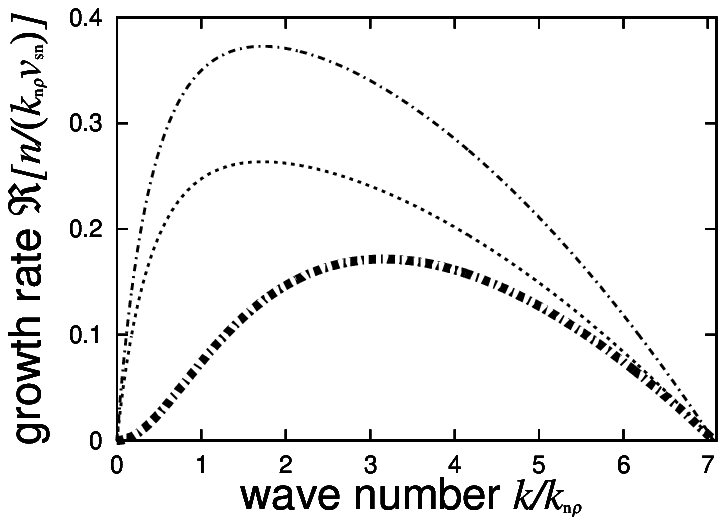}
	\caption
	{
	Growth rates of the condensation mode for the instability
	characterized by $\mu_{\rm n}$ (the thick dot-dashed curve), 
	when $ v_{\rm A} / v_{\rm sn} = 600$ and 
	$\nu _{{\rm ni}0} / ( k_{\rm n \rho} v_{\rm sn} ) = 5$. 
	The mode of $\mu_{\rm i}$ is completely stabilized 
	by the magnetic field and does not appear in this figure.
	The horizontal axis is the normalized wave number $ k / k_{\rm n \rho} $.
    The vertical axis corresponds to the normalized growth rate 
    $ \Re [n / ( k_{\rm n \rho} v_{\rm sn} ) ]$.
    The difference between Figure \ref{B600C03} and Figure \ref{B600C5}
     is the friction strength. 
    For comparison, there are also displayed
     condensation modes of the ion component (the thin dot-dashed curve) 
	and the neutral component (the thin dashed curve) when $ B=0$ and $\nu _{{\rm ni}0} = 0$.
    }
	\label{B600C5}
\end{figure}

In Figure \ref{B600C5}, the friction is stronger than that in Figure
 \ref{B600C03}, but the magnetic field is the same [i.e., 
$ v_{\rm A} / v_{\rm sn} = 600$, 
$\nu _{{\rm ni}0} / ( k_{\rm n \rho} v_{\rm sn} ) =5$]. 
For comparison,
the two thin curves are displayed like Figure \ref{B0C003}.
A mode characterized by $\mu_{\rm i}$ is completely stabilized,
and then it disappears here. 
A mode characterized by $\mu_{\rm n}$ is still
unstable, while the growth rate is reduced.
That is,
even in weakly ionized fluid, the growth of the thermal
instability is affected by the magnetic field through the friction.
This decrement of the growth rate is significant at longer scale.
We also confirm that the characteristic wavelength at the maximum growth
rate of the instability shifts shortward 
as the growth rate diminishes.
This occurs since the friction is effective at larger scale 
and the conduction is effective at smaller scale, 
so that the growth rate at larger scale decreases 
significantly. Then,
the characteristic wavelength at the maximum growth rate becomes small.

It should be emphasized that all the critical wavelengths of the unstable
mode characterized by $\mu_{\rm n}$ are equivalent among Figures
\ref{B0C0} -- \ref{B600C5}. 
The frictional terms in the right hand side of the dispersion relation (\ref{dis-rela}) are independent of the critical wavelength for the condensation mode, as mentioned in subsection \ref{subsec:results:critical_wavelength}.
The critical wavelength of the mode
relating $\mu_{\rm n}$
is not changed by the magnetic field with the friction as long as
the thermal conduction is not changed by the magnetic field. 
This point is investigated more in the following subsection \ref{subsec:critical_wavelength_conduction}.


\section{DISCUSSION}
\label{S_Discuss}

We study the basic property of the thermal instability 
of the weakly ionized plasma in the previous sections. 
In this section,
basing on our understanding the thermal instability,
we shall discuss the structure formation.

\subsection{Critical Wavelength and Thermal Conduction}
\label{subsec:critical_wavelength_conduction}

First in this section, the critical wavelength of the weakly ionized plasma in the magnetic field is discussed. 
As mentioned in section \ref{S_Results}, 
the critical wavelength of the mode relating $\mu_{\rm n}$
is not changed by the magnetic field with the friction as long as
the thermal conduction is not changed by the magnetic field. 
In fact, the thermal conductivity is varied owing to the magnetic
field. 
In this meaning, we may oversimplify the problem in this paper.

However, the thermal conductivity will decrease as the magnetic field strength increases, and then the critical length become smaller, according to equation (\ref{eq:appendix:criticallength:neutral}). 
Thermal conduction erases the temperature perturbation
and thermally stabilizes the system, especially at smaller scale. 
The larger thermal conduction makes the critical wavelength longer, owing to
the criterion \ref{crite_cond} or \ref{crite_ion}, according to \citet{F65}.
On the other hand, the magnetic field lessens the thermal conduction
of the ion component, because the ion component winds around a magnetic
field.

Therefore, we can insist that the critical length for the thermal instability of the weakly ionized plasma can not be enlarged by the friction and the magnetic field at least, 
while the effect of the magnetic field can influence not only the ion but the neutral component via the ion-neutral friction.
This statement has universal validity for the thermal instability of the weakly ionized plasma. 
The effect of the friction and the magnetic field is mainly reduction of the
growth rate of the thermal instability. 
On the other hand, the magnetic field directly enlarges the critical wavelength
of the mode characterized by $\mu_{\rm i}$ as expected in the
one fluid plasma model.

\subsection{Spatial Distribution of Magnetized Interstellar Medium}

In the previous section, we learn that the suppression of
the thermal instability is characterized by the ion-neutral
friction, strength of the magnetic field, and the direction 
of the field. Interestingly, the suppression of the instability
is different owing to the magnetic field direction. This suggests
the anisotropic nature of the suppression of the instability
is imprinted in the morphology of the structure.
In this standing
point, we shall discuss how the structure formation proceeds
in the magnetized ISM.

\subsubsection{Growth Rate Parallel to the Magnetic Field}

In an usual one fluid approach,
the motion of the plasma parallel to the magnetic
field is regarded not to be affected. However, Figure \ref{B0C003}
suggests that the growth rate of the thermal instability at larger scale 
decreases via the ion-neutral friction. This suppression of the instability
yields the delay of the evolution in comparison to the evolution
under the MHD approximation without the ion-neutral friction.

\subsubsection{Distribution of Partially Ionized Plasma 
Perpendicular to the Magnetic Field Line}
\label{sss_spread_of_part}

Here, we suggest that the spatial distribution of the partially ionized
plasma is elongated, whose semi-major axis is vertical to the
magnetic field line. 
Figures \ref{B0C003} and \ref{B06C003},
both of which friction strengths are the same, show the feasibility.
The thermal instability in Figure \ref{B0C003} corresponds to the mode along
the magnetic field line since $\mbox{\boldmath$B$} = 0$, while that in
Figure \ref{B06C003} is vertical to the magnetic field line. The growth
of thermal instability vertical to the line is suppressed by both of the
magnetic field and the friction, while the growth along the line is
suppressed only by the friction. Thus, the growth rate along the
magnetic field line is larger than that vertical to the line, and then
the resultant morphology of the structure via the thermal instability 
is elongated.

\subsubsection{Difference of Spatial Distribution of Ion 
and Neutral Components}
\label{ssS_Difference_in_Spread}

We insist the growth rate depending on the direction of the field
results in the morphology of interstellar structure. 
Furthermore,
we learn in \S3 the mode characterized by $\mu_{\rm i}$
are stabilized by the friction and field,
while that by $\mu_{\rm n}$ can grow as found 
in Figure \ref{B06C003}.
Figure \ref{B600C5} shows that although the mode characterized by 
$\mu_{\rm i}$ is completely stabilized, 
the mode characterized by $\mu_{\rm n}$ is still unstable.
Consequently, we can insist that 
the neutral component condenses owing to the thermal
instability, while the ion component still keeps spreading.

It is noticed that the ion component may condense
since it is dragged by the
neutral component contracting owing to the thermal instability.
However, once the distribution of the ion and the neutral components
separates and the motion of the two becomes independent there, the ion
component in the area is 
thermally stable owing to the magnetic field. This means 
the magnetic pressure lets the ion component be spread in the area
even if the neutral component condenses.

This thin ion component left
behind by the contracting neutral component will be observed as a halo
around H {\rm \textsc{i}} clouds, and it emits, especially some forbidden
lines. This halo possibly has wide distribution vertical to the
magnetic field, as discussed in subsubsection \ref{sss_spread_of_part}.
This is because the ion component can condense, parallel to the magnetic field like the neutral.
The magnetic pressure
can not prevent the motion of the ion component efficiently
as found in Figure \ref{B0C003}. Thus, the morphology of the ion halo 
is also elongated according to the orientation of the field line.

\subsubsection{Origin of the Small Clumpinesses of Partially Ionized Plasma}
\label{subsubsec:small_clumpinesses_of_partial}

It is suggested that 
there are small clumpy structures of ISM.
Here, we examine whether the clumpiness originates
via the thermal instability. 
According to Figures \ref{B600C03} and
\ref{B600C5}, the growth rate of the thermal instability of partially
ionized plasma depends on the strength of the friction. The scale
length at the maximum growth rate of the instability becomes smaller,
owing to the suppression via the friction which is effective at larger
scale. Thus, we can expect the small clumpiness appears
selectively via the thermal instability since the ion-neutral friction 
suppresses the large scale condensations.

If we are interested in the small clumpiness of ISM,
we should mention about the thermal conduction in relation to the
magnetic field. 
According to subsection \ref{subsec:critical_wavelength_conduction},
 the small clumpiness may the results of the decrement
of the conduction. 
It is noticed that the conduction along the field line
is not so affected, and then there is still ambiguity
for the reduced conduction coefficient to perform the key role
for the origin of the small clumpiness.

It is also noticed that 
the thermal instability is a slow process relative to the
dynamical processes. 
If molecular clouds form from the H {\rm \textsc{i}} medium via the gravitational
instability, the small clumpiness inside the molecular clouds
appears after the formation of the clouds. 
This trend is enhanced if the effects of the magnetic field
and friction are concerned. 
This is because the friction and the field suppress the large scale growth,
while the short wavelength modes are still unstable.
In a enough long
span for the growth, the tiny structures can appear in ISM and affect
the evolution of its structure especially as the non-linear effects.
Thus, if we want to know 
how faint and small structures come to be observable,
the growth time-scale of the thermal instability 
should be precisely examined. At this time, we never forget 
the effect of the magnetic field and the ion-neutral friction.

\subsection{Ineffectiveness of the Magnetic Field}
\label{S_Discuss_Satu}


The modes in Figures \ref{B600C03} and \ref{B6000C03} have similar growth rate, while the magnetic fields are significantly different in each other. 
It implies that the suppression of the growth by increasing the magnetic field with the fixed friction has limits.
By the way, 
a dispersion relation of only ionized fluid in a magnetic field, which
is derived by equalizing the second bracket on the left-hand side in
equation (\ref{dis-rela}) to be zero, is reduced to
\begin{equation}
n^3 
+ n^2 v_{\rm si} k_{\rm iT}
= 0
,
\end{equation}
when $k$ approaches $0$.
This is solved to be 
\begin{equation}
n=0
~
{\rm or}~
n = -v_{\rm si} k_{\rm iT} .
\end{equation}
Thus, we find that this solution is independent of a magnetic field. 
Even if the magnetic
field becomes strong, the value of the growth rate at $k \sim 0$ is
scarcely changed,
 where the friction is more effective than at small wavelength. 
Therefore, once the effect of the friction is fixed, 
increasing the magnetic field strength beyond a threshold level apparently becomes ineffective for the suppression of the growth.
Not only information of the magnetic field but the friction is necessary to investigate the evolution of the thermal instability of the weakly ionized plasma.


\subsection{Comment on One Fluid Approximation}

The formulation of the fully one component plasma
can not be always applicable to study 
the thermal instability of weekly ionized
plasma. 
This is because there is the growing mode even when there exist the
magnetic field and friction, whose growth rate is comparable to that of
the neutral component when there are no magnetic field and friction.
Indeed, this trend is found from the comparison among
Figures \ref{B06C003}, \ref{B600C03} and \ref{B6000C03}.

In addition, regarding weakly ionized plasma including 
the friction effect, the
friction does not change the critical wavelength $\lambda_{\rm F}$. 
The
friction tends to efficiently reduce the growth at much larger scale
than the $\lambda_{\rm F}$. 
This feature yields that the critical
wavelength in the weakly ionized plasma is not changed by the magnetic
field via the friction, as mentioned in section \ref{S_Results} and subsection \ref{subsec:critical_wavelength_conduction}. 
Even if the magnetic field becomes strong, only the
growth rate is reduced, keeping the critical wavelength constant. 
This is also the difference from the simple analysis of the one component
magnetized plasma. The previous study of the one component plasma
shows the stabilization at the smaller scales, while 
the current study insist $\lambda_{\rm F}$ is unchanged
even if the effect of the field is concerned.


\section{Application to Typical H {\rm \textsc{i}} Region}
\label{Application_to}

In this section, we shall study the origin of the
small scale structure of H {\rm \textsc{i}} cloud as an example.
The thermal instability of the typical H {\rm \textsc{i}} region with
C {\rm \textsc{ii}} cooling is briefly discussed here.

\subsection{Assumptions}

We assume that the temperature of H {\rm \textsc{i}} is $100$ K. 
The fluid mainly consists of H atom, and then $\mu_{n}=1$, $\mu_{i}=1/2$. 
We assume that $K = 4.9 \times 10^3$ erg cm$^{-1}$ s$^{-1}$ K$^{-1}$, consulting the result of \citet{Ul70} as thermal conductivity.
In addition, we adopt the collision cross-section as $\langle
\sigma v \rangle = 2 \times 10^{-9}$ cm$^3$ s$^{-1}$ for the typical friction value according to \citet{Na76}. 
We assume that interstellar pressure is about $10^{-12}$ erg cm$^{-3}$. 
Then, if the ISM is in equilibrium under the pressure
of $10^{-12}$ erg cm$^{-3}$, the number density of the neutral
component $n _{\rm neutral}$ and the magnetic field strength $B$ are
about $71.9$ cm$^{-3}$ and about $10^{-6}$ Gauss, respectively. The
adiabatic $\gamma$ is assumed to be $5/3$.

According to \citet[Fig. 3 (b)]{Wo95}, C {\rm \textsc{ii}} cooling is
dominant when $T = 100$ K. 
Then we adopt the expression for the C {\rm \textsc{ii}} cooling according to \citet[their Appendix B]{Wo95}.
Assuming that the fluid consists of H atom, whose mass is expressed as
m$_{\rm H}$, the cooling function $\Lambda$ is
\begin{equation}
\Lambda
=
\frac{1}{m_H} \cdot
2.54 \times 10^{-14} A_{\rm C} f_{\rm CII} \cdot
[\gamma ^{\rm H ^0} n _{\rm H^0 } + \gamma ^{\rm e} n_{\rm e} ] \cdot
\exp \left[- \frac {92} {T} \right] ~~ {\rm ergs ~ s}^{-1} ~ {\rm g} ^{-1}
.
\end{equation}
Here, $n_{\rm X}$ is the density of element X.
$\gamma^{\rm H^0}$ and $\gamma ^{\rm e}$ are the collisional de-excitation rate
coefficients of C {\rm \textsc{ii}} with neutral hydrogen
and electrons, respectively. 
The constant $f_{\rm CII}$ is the fraction
of C {\rm \textsc{i}} in C {\rm \textsc{ii}}. 
$A_{\rm C}$ is the relative abundance of element C, defined as $n_{\rm C} / n $. 
$n$ is the density of hydrogen nuclei, defined as $n = n_{\rm H^+} + n_{\rm H^0}$.
According to \citet{Wo95},
\begin{equation}
\gamma ^{\rm H^0} = 8.86 \times 10^{-10}
{\rm cm} ^3 {\rm s} ^{-1},
\end{equation}
\begin{equation}
\gamma ^{\rm e}
=
2.1 \times 10^{-7}
\left( \frac{T}{10^2} \right) ^{-0.5}
\left[
1.80 + 0.484 \left( \frac{T}{10^4} \right)
 +4.01 \left( \frac{T}{10^4} \right) ^2
  - 3.39 \left( \frac{T}{10^4} \right) ^3
  \right],
\end{equation}
and $A_{\rm C} = 3\times 10^{-4}$.

The ionization degree and the fraction of C {\rm \textsc{i}} in C {\rm \textsc{ii}}, i.e., $f_{\rm CII}$ are ambiguous. 
Then, the ionization degree $\chi$ and the constant $f_{\rm CII}$ are hypothesized in some
cases. We shall investigate the instability when $f_{\rm CII} =
10^{-2}$ and $f_{\rm CII} = 10^{-4}$. 
The more the amount of C {\rm \textsc{ii}} increases, the more efficient the cooling becomes. 
The ionization degree $\chi$ is assumed to be $10^{-6}$ according to the
results around $n_{\rm H} \sim 100$ cm$^{-3}$ by \citet[Fig.1]{KaNi00}.
We also investigate the instability on the extreme case when $\chi =
10^{-2}$. In addition, we assume that
$\Lambda _{\rm n \rho} = \Lambda _{\rm i \rho}$,
$\Lambda _{\rm n T} = \Lambda _{\rm i T}$,
and
$K_{{\rm n}0} = K_{{\rm i}0}$.
This is because the ionization degree is very small and the mixing of
the neutral and the ion components is sufficient. Then the temperatures
$T_{\rm n}$ for the neutral and $T_{\rm i}$ for the ion become
equivalent much faster than the cooling time. For example, if the
temperature of the system changes by $\Delta T$ during the slight time
$\Delta t$, the energy changes for the neutral and the ion are expressed
as,
\begin{displaymath}
n_{\rm n} k_{\rm B} \Delta T \sim \rho_{\rm n} \Lambda _{\rm n} \Delta t
\end{displaymath}
and
\begin{displaymath}
n_{\rm i} k_{\rm B} \Delta T \sim \rho_{\rm i} \Lambda _{\rm i} \Delta t.
\end{displaymath}
From these equations,
\begin{displaymath}
\frac{\rho_{\rm n}}{n_{\rm n}} \Lambda _{\rm n} \sim
\frac{\rho_{\rm i}}{n_{\rm i}} \Lambda _{\rm i}
.
\end{displaymath}
Therefore, owing to $m_{\rm n} \sim m_{\rm i}$ for H atoms,
it is derived that
$
\Lambda _{\rm n} \sim \Lambda _{\rm i}
$.

\subsection{Results and Implications}

Figure \ref{100K_kai10_-6_f_10-2_1} shows the growth rate 
of the thermal instability when $\chi =
10^{-6}$ and $f_{\rm CII} = 10^{-2}$. The critical wavelength for the
instability is estimated to be about $0.0135$ pc owing to the criterion
(\ref{crite_cond}). 
	It is noted that 
	$\alpha _{\rm n}=0.920$, $\alpha _{\rm i}=920000$, 
	$\beta _{\rm n}=2.19 \cdot 10^{-6}$, 
	$\beta _{\rm i}=2.73 \cdot 10^{-7}$, 
	$\kappa _{v _{\rm A}}=219$, $C_{\nu}=0.775$, 
	$\kappa _{\rho}= \chi / \mu_{\rm ni}^2 = 5 \cdot 10^{-7}$, 
	according to our notation summarized in Appendix \ref{Ap_nor}.
Figure \ref{100K_kai10_-6_f_10-4_1} shows the
growth rate when $\chi = 10^{-6}$ and $f_{\rm CII} = 10^{-4}$. 
The critical wavelength for the instability is estimated to be about $0.135$
pc owing to the criterion (\ref{crite_cond}).
Here, 
	$\alpha _{\rm n}=0.920$, $\alpha _{\rm i}=920000$, 
	$\beta _{\rm n}=2.19 \cdot 10^{-8}$, 
	$\beta _{\rm i}=2.73 \cdot 10^{-9}$, 
	$\kappa _{v _{\rm A}}=219$, $C_{\nu}=77.5$, 
	$\kappa _{\rho}= \chi / \mu_{\rm ni}^2 = 5 \cdot 10^{-7}$.
Figure \ref{100K_kai10_-2_f_10-2_1} shows the growth rate when $\chi = 10^{-2}$
and $f_{\rm CII} = 10^{-2}$. The critical wavelength for the
instability is estimated to be about $0.00239$ pc owing to the criterion
(\ref{crite_cond}).
We adopt 
	$\alpha _{\rm n}=0.517$, $\alpha _{\rm i}=51.7$, 
	$\beta _{\rm n}=1.15 \cdot 10^{-5}$, 
	$\beta _{\rm i}=1.44 \cdot 10^{-6}$, 
	$\kappa _{v _{\rm A}}=2.19$, $C_{\nu}=1470$, 
	$\kappa _{\rho}= \chi / \mu_{\rm ni}^2 = 5 \cdot 10^{-3}$.
To understand more, in each set of these figures, 
the growth rates of the following three cases 
are also plotted: 
(1) the magnetic field is hundred times as the typical value, 
(2) the friction is hundred times as the typical value, 
and (3) both of the magnetic field and the friction are hundred
times as the typical value.
It is noticed
that the value of the horizontal and the vertical axes is normalized by
$k_{{\rm n} \rho}$, which is proportional to $f_{\rm CII}$.
The more C {\rm \textsc{ii}} (i.e., $f_{\rm CII}$) increases, the shorter the
critical wavelength becomes and the more the growth rate increases.

The each panel (b) in Figures \ref{100K_kai10_-6_f_10-2_1},
\ref{100K_kai10_-6_f_10-4_1}, and \ref{100K_kai10_-2_f_10-2_1} shows
characteristic results. 
In these panels, we find some characteristic feature of the example.
At larger wavenumber than $k_{\rm max}$, which corresponds
to the maximum growth rate, the instability grows
up as if there is no friction and field. 
We insist that the critical wavelengths of these cases are not enlarged because the neutral component does not
suffer the effect of the magnetic field directly.
At smaller wavenumber than $k_{\rm max}$, 
the growth rate is reduced significantly. 
That is, the growth of the instability in the significantly large-scale 
is suppressed. This may indicate the quasistatic evolution 
like the gravitational contraction via the so-called ambipolar diffusion,
if the condensation due to the thermal instability is balanced
by the magnetic field via the friction.

Except the cases of $\chi = 10^{-2}$, 
the overall trend of the results is
the same as mentioned in section \ref{S_Results}. The instability in
the very low ionization degree cases of $\chi = 10^{-6}$ 
is hardly suppressed. 
In Figures \ref{100K_kai10_-6_f_10-2_1} and \ref{100K_kai10_-6_f_10-4_1}, the growth at only the large scale can be suppressed by the even large friction and magnetic field. 
In the case of $\chi = 10^{-2}$ (Figure \ref{100K_kai10_-2_f_10-2_1}), the ionization degree is higher than
the other cases, then the ion-neutral friction is more efficient.
The growth rate in the panel (d) of Figure
\ref{100K_kai10_-2_f_10-2_1} is reduced all over wavelength because
of the large friction. 
The growth of the mode in the panel (e) in Figure \ref{100K_kai10_-2_f_10-2_1} is extremely suppressed all over the wavelength because of the large friction and the magnetic field.
Especially, the growth rate at the significantly large scale (i.e., near the origin of wavenumber) approaches to the zero owing to the friction. 
This figure shows
the possibility that the thermal instability is suppressed as well as in
Figure \ref{B600C5}. 
The magnetic field can become strong for the ion
component to be frozen in,
 and the friction between the ion and the neutral components can be strong even for the high density, which are expected at
the final stage of the formation.

It is noticed that under the conditions adopted in this section, 
the neutral
component can always satisfy the thermal instability criterion,
while the ion component does not satisfy the criterion.
Importantly, the ionization degree is very small and then the ion component
is too rare although the cooling 
owing to the ion component is effective 
for the thermal instability.
Then, the thermal instability of the partially ionized plasma can not be
suppressed except for the extreme case such as in the panel (e) of
Figure \ref{100K_kai10_-2_f_10-2_1}. 
The partially ionized plasma
can start to contract via the thermal instability in typical H {\rm \textsc{i}} region, even when there exists a magnetic field. 
Even if the magnetic field becomes large, the growth rate is
hardly suppressed except for the extreme case.

\section{SUMMARY}
\label{S_Summary}

The thermal instability of the weakly ionized fluid is 
investigated with a linear perturbation analysis. 
The plasma is assumed to consist of two fluids of 
the ion and the neutral components.
With this approach, 
the effect of the friction between the ion and the neutral components 
and the magnetic field are important.
Here, the properties of the thermal instability of weakly ionized plasma 
and the observational implications are summarized.

\begin{enumerate}

\item
Modes relating to $\mu_{\rm n}$ are not stabilized by the magnetic field.
This is because the neutral component feels the field via the ion-neutral
friction.

\item
Modes relating to $\mu_{\rm i}$ are directly stabilized by the magnetic 
field.
This means that when the magnetic field is large, 
the ion component is hard to follow the motion of the neutrals
which condensate owing to the thermal instability.

\item
The growth rate of the mode vertical to the magnetic field is reduced
 by the magnetic field and the ion-neutral friction.

\item
The growth rate of the mode along the magnetic field
decreases owing to the ion-neutral friction, 
especially at large scale. 
The suppression of the instability is more ineffective 
than that of the mode vertical to the magnetic field. 

\item
The critical wavelength for the thermal instability is 
not affected by the friction.
The effect owing to the two fluid approximation is 
the reduction of the growth rate of the thermal instability.
The magnetic field makes the critical wavelength of modes relating 
$\mu_{\rm i}$ larger. 

\item
To study the thermal instability of the partially ionized plasma,
one fluid approximation is not always useful.

\item 
The ion-neutral friction and the magnetic field affect 
the distribution or morphology of ISM, 
especially after the long time compared to the free-fall time.
The difference between the growth rate along and perpendicular 
to the magnetic field is important. 
The partially ionized plasma possibly is elongated perpendicular 
to the magnetic field.
In addition, the neutral component and the ion component of 
weakly ionized fluid
in the magnetic field are possibly separated each other. 
Then, the neutral component condenses owing to thermal instability,
while the ion component left behind by the contracting neutral 
component still keeps spreading. 
Therefore, it is possibly observed that the weakly ionized fluid 
is elongated vertically to the magnetic field and surrounded 
by a halo which is rare and diffuse ions and emits forbidden lines.

\end{enumerate}

To conclude, 
the treating both of the independent motions of the neutral 
and the ion components yield the ion-neutral friction. 
The friction is important for the thermal instability of weakly 
ionized fluid, especially when studying the structure formation 
such as molecular cloud at its final stage of formation. 
Our study indicates that the fully ionized plasma approximation 
or totally neutral fluid approximation is not always applicable 
to weakly ionized plasma.

\acknowledgements

We would like to thank the referee for constructive comments.
We would like to thank Hiroshi Koyama for useful discussions.
We would like to thank Tsuyoshi Inoue for useful comments.
We would like to thank Jun Fukue for useful comments and encouragements.
We would like to thank Tetsuya Nagata for encouragements and helps.
Tsubasa Fukue is supported by Research Fellowships of the Japan Society for the Promotion
of Science for Young Scientists.
This work was partially supported by KAKENHI 18$\cdot$3219.
This work is partially supported by the Grant-in-Aid for the 21st Century COE "Center for Diversity and Universality in Physics" from the Ministry of Education, Culture, Sports, Science and Technology (MEXT) of Japan.

\appendix
\section*{Appendix}

\section{Derivation of the Dispersion Relation by Linear Perturbation Analysis}
\label{Ap_del}

We derive the dispersion relation (\ref{dis-rela}) in this Appendix.
We adopt the perturbations as equation (\ref{perturb}), 
that is, 
$
a(\mbox{\boldmath$r$}, t) = a_{1} 
\exp (nt + i \mbox{\boldmath$k$} \cdot \mbox{\boldmath$r$})
$.
Basic equations (\ref{cont_n})--(\ref{inductioneq}) are then linearized as follows: 
\begin{eqnarray}
\rho_{{\rm n} 1} n 
+ i \rho_{{\rm n} 0} \mbox{\boldmath$v$}_{{\rm n} 1} \cdot \mbox{\boldmath$k$}
&=& 0
,\\
n \rho_{{\rm n} 0} \mbox{\boldmath$v$}_{{\rm n} 1} 
&=&
- p_{{\rm n} 1} i \mbox{\boldmath$k$} 
- \rho_{{\rm n} 0} \nu_{\rm ni} (\mbox{\boldmath$v$}_{{\rm n} 1} - \mbox{\boldmath$v$}_{{\rm i} 1} )
,\\
\frac{1}{\gamma -1} p_{{\rm n} 1} n 
- \frac{\gamma}{\gamma -1} \frac{p_{{\rm n} 0}}{\rho_{{\rm n} 0}} n \rho _{{\rm n} 1} 
&=&
- \rho _{{\rm n} 0} (\rho _{{\rm n} 1} \Lambda _{{\rm n} \rho} + T_{{\rm n} 1} \Lambda _{{\rm n} T}) 
- K_{{\rm n} 0} T_{{\rm n} 1} k^2
,\\
\frac{p_{{\rm n} 1}}{p_{{\rm n} 0}}
&=& \frac{\rho_{{\rm n} 1}}{\rho_{{\rm n} 0}}
+ \frac{T_{{\rm n} 1}}{T_{{\rm n} 0}}
,\\
\rho_{{\rm i} 1} n 
+ i \rho_{{\rm i} 0} \mbox{\boldmath$v$}_{{\rm i} 1} \cdot \mbox{\boldmath$k$}
&=& 0
,\\
n \rho_{{\rm i} 0} \mbox{\boldmath$v$}_{{\rm i} 1} 
&=&
- p_{{\rm i} 1} i \mbox{\boldmath$k$} 
- \rho_{{\rm n} 0} \nu_{\rm ni} (\mbox{\boldmath$v$}_{{\rm i} 1} - \mbox{\boldmath$v$}_{{\rm n} 1} ) 
+ \frac{1}{4 \pi} i (\mbox{\boldmath$k$} \times \mbox{\boldmath$B$}_{1}) \times \mbox{\boldmath$B$}_{0}
,\\
\frac{1}{\gamma -1} p_{{\rm i} 1} n 
- \frac{\gamma}{\gamma -1} \frac{p_{{\rm i} 0}}{\rho_{{\rm i} 0}} n \rho _{{\rm i} 1} 
&=& 
- \rho _{{\rm i} 0} (\rho _{{\rm i} 1} \Lambda _{{\rm i} \rho} + T_{{\rm i} 1} \Lambda _{{\rm i} T}) 
- K_{{\rm i} 0} T_{{\rm i} 1} k^2
,\\
\frac{p_{{\rm i} 1}}{p_{{\rm i} 0}}
&=& \frac{\rho_{{\rm i} 1}}{\rho_{{\rm i} 0}}
+ \frac{T_{{\rm i} 1}}{T_{{\rm i} 0}}
,
\end{eqnarray}
and
\begin{eqnarray}
n \mbox{\boldmath$B$}_{1} 
&=& i [ \mbox{\boldmath$v$}_{{\rm i} 1} (\mbox{\boldmath$B$}_{0} \cdot  \mbox{\boldmath$k$}) 
- \mbox{\boldmath$B$}_{0} (\mbox{\boldmath$v$}_{{\rm i} 1} \cdot \mbox{\boldmath$k$}) ]
.
\label{linear_eq_induc}
\end{eqnarray}

Using conditions 
$
\mbox{\boldmath$B$}_{0}
\perp
\mbox{\boldmath$k$}
$,
$
\mbox{\boldmath$k$} \parallel
\mbox{\boldmath$v$}_{{\rm i} 1}
$,
$
\mbox{\boldmath$k$} \parallel
\mbox{\boldmath$v$}_{{\rm n} 1}
$,
and 
$
\mbox{\boldmath$B$}_{0} \parallel
\mbox{\boldmath$B$}_{1}
$, 
this multi-dimensional issue is reduced to one dimensional issue.
Then, vanishing of the determinant of coefficients, a dispersion relation for partially ionized plasma is derived as follows: 
\clearpage
\begin{eqnarray}
\lefteqn{
\left[
n^3 
+ n^2 \left(v_{\rm sn} k_{\rm nT} + v_{\rm sn} k \frac{k}{k_{\rm n K}} \right)
+ n (v_{\rm sn} k)^2 
+ \frac{1}{\gamma} (v_{\rm sn} k)^2
\left( v_{\rm sn} k_{\rm nT} - v_{\rm sn} k_{\rm n \rho}
+ v_{\rm sn} k \frac{k}{k_{\rm n K}} \right)  
\right] 
}
\nonumber \\
& & {}
\lefteqn{
\times 
\left[
n^3 
+ n^2 \left(v_{\rm si} k_{\rm iT} + v_{\rm si} k \frac{k}{k_{\rm i K}} \right)
+ n (v_{\rm si} k)^2 
+ n (v_{\rm A} k)^2
\right.
}
\nonumber \\
& & {} 
\hspace{1.1cm}
\lefteqn{\left.
+ \frac{1}{\gamma}
\left( v_{\rm si} k_{\rm iT} + v_{\rm si} k \frac{k}{k_{\rm i K}} \right)
\left( v_{\rm si}^2 k^2 + \gamma v_{\rm A}^2 k^2 \right)
- \frac{1}{\gamma} (v_{\rm si} k)^2 (v_{\rm si} k_{\rm i \rho})
\right]
}
\nonumber \\
& = &
-\nu _{{\rm ni} 0} \frac{\rho _{{\rm i} 0} + \rho _{{\rm n} 0}}{\rho _{{\rm i} 0}} \cdot
\left \{
n^5
+ n^4 
\left( 
v_{\rm si} k_{\rm iT} + v_{\rm si} k \frac{k}{k_{\rm i K}} 
+ v_{\rm sn} k_{\rm nT} + v_{\rm sn} k \frac{k}{k_{\rm n K}} 
\right)
\right.
\nonumber \\
& & {}
+ n^3 
\left[ 
v_{\rm sn} k \frac{k}{k_{\rm nK}} 
v_{\rm si} k \frac{k}{k_{\rm iK}}
+ v_{\rm sn} k \frac{k}{k_{\rm nK}} v_{\rm si} k_{\rm iT}
+ v_{\rm si} k \frac{k}{k_{\rm iK}} v_{\rm sn} k_{\rm nT}
\right.
\nonumber \\
& & {}
\hspace{1.1cm}
\left. 
+ v_{\rm sn} k_{\rm nT} v_{\rm si} k_{\rm iT}
+ k^2 \left( \gamma \frac{p_{{\rm i} 0} + p_{{\rm n} 0}}{\rho _{{\rm i} 0} + \rho _{{\rm n} 0}} + \frac{B_{0}^2}{4 \pi} \frac{1}{\rho _{{\rm i} 0} + \rho _{{\rm n} 0}} \right)
\right]
\nonumber \\
& & {}
+ n^2 
\left[
\frac{p_{{\rm n} 0} + p_{{\rm i} 0} \gamma}{\rho _{{\rm i} 0} + \rho _{{\rm n} 0}} k^2 
(v_{\rm sn} k_{\rm nT} + v_{\rm sn} k \frac{k}{k_{\rm n K}})
- \frac{p_{{\rm n} 0}}{\rho _{{\rm i} 0} + \rho _{{\rm n} 0}} k^2 v_{\rm sn} k_{\rm n \rho}
\right.
\nonumber \\
& & {}
\hspace{1.1cm}
\left.
+ \frac{p_{{\rm i} 0} + p_{{\rm n} 0} \gamma}{\rho _{{\rm i} 0} + \rho _{{\rm n} 0}} k^2 
(v_{\rm si} k_{\rm iT} + v_{\rm si} k \frac{k}{k_{\rm i K}})
- \frac{p_{{\rm i} 0}}{\rho _{{\rm i} 0} + \rho _{{\rm n} 0}} k^2 v_{\rm si} k_{\rm i \rho}
\right.
\nonumber \\
& & {}
\hspace{1.1cm}
\left.
+ \frac{B_{0}^2}{4 \pi} \frac{1}{\rho _{{\rm i} 0} + \rho _{{\rm n} 0}} 
\left( 
v_{\rm si} k_{\rm iT} + v_{\rm si} k \frac{k}{k_{\rm i K}} 
+ v_{\rm sn} k_{\rm nT} + v_{\rm sn} k \frac{k}{k_{\rm n K}} 
\right)
\right]
\nonumber \\
& & {} 
+ n 
\left[
\left( \frac{B_{0}^2}{4 \pi} \frac{1}{\rho _{{\rm i} 0} + \rho _{{\rm n} 0}} k^2 + \frac{p_{{\rm i} 0} 
+ p_{{\rm n} 0}}{\rho _{{\rm i} 0} + \rho _{{\rm n} 0}} k^2 \right)
\right.
\nonumber \\
& & {}
\hspace{1.1cm}
\left.
\cdot
\left(
v_{\rm sn} k \frac{k}{k_{\rm n K}} v_{\rm si} k \frac{k}{k_{\rm i K}}
+ v_{\rm sn} k \frac{k}{k_{\rm n K}} v_{\rm si} k_{\rm iT} 
+ v_{\rm si} k \frac{k}{k_{\rm i K}} v_{\rm sn} k_{\rm nT}
+ v_{\rm sn} k_{\rm nT} v_{\rm si} k_{\rm iT}
\right)
\right.
\nonumber \\
& & {}
\hspace{1.1cm}
\left.
- \frac{p_{{\rm i} 0}}{\rho _{{\rm i} 0} + \rho _{{\rm n} 0}} k^2 v_{\rm si} k_{\rm i \rho}
(v_{\rm sn} k_{\rm nT} + v_{\rm sn} k \frac{k}{k_{\rm n K}})
\right.
\nonumber \\
& & {}
\hspace{1.1cm}
\left.
\left.
- \frac{p_{{\rm n} 0}}{\rho _{{\rm i} 0} + \rho _{{\rm n} 0}} k^2 v_{\rm sn} k_{\rm n \rho}
( v_{\rm si} k_{\rm iT} + v_{\rm si} k \frac{k}{k_{\rm i K}} )
\right]
\right\}
.
\label{dis-rela_partial}
\end{eqnarray}

We emphasize that this dispersion relation is not yet assumed to be weakly ionized plasma, such as $\rho _{{\rm i}0} \ll \rho _{{\rm n}0}$.
Here are
\begin{displaymath}
v_{\rm sn}^2 = \gamma \frac{p_{{\rm n}0} } { \rho_{{\rm n}0} }, ~~
v_{\rm si}^2 = \gamma \frac{p_{{\rm i}0} } { \rho_{{\rm i}0} }, ~~ 
v_{\rm A}^2 = \frac{B _0 ^2}{4 \pi \rho _{{\rm i}0} },
\end{displaymath}

\begin{displaymath}
k_{\rm n \rho}=
\frac{\mu_{\rm n}(\gamma -1) \rho_{{\rm n}0} \Lambda _{\rm n \rho} }
{{\cal R} v_{\rm s n} T_{{\rm n} 0}}, ~~
k_{\rm n T}=
\frac{\mu_{\rm n}(\gamma -1) \Lambda _{\rm n T} }
{{\cal R} v_{\rm s n}}, ~~
k_{\rm n K}=
\frac{{\cal R} v_{\rm s n} \rho _{{\rm n} 0}}
{\mu_{\rm n}(\gamma -1) K_{{\rm n}0}},
\end{displaymath}

\begin{displaymath}
k_{\rm i \rho}=
\frac{\mu_{\rm i}(\gamma -1) \rho_{{\rm i}0} \Lambda _{\rm i \rho} }
{{\cal R} v_{\rm s i} T_{{\rm i} 0}}, ~~
k_{\rm i T}=
\frac{\mu_{\rm i}(\gamma -1) \Lambda _{\rm i T} }
{{\cal R} v_{\rm s i}}, ~~
k_{\rm i K}=
\frac{{\cal R} v_{\rm s i} \rho _{{\rm i} 0}}
{\mu_{\rm i}(\gamma -1) K_{{\rm i}0}},
\end{displaymath}

\begin{displaymath}
\Lambda _{\rm n \rho} \equiv 
\left( \frac{\partial \Lambda _{\rm n}}{\partial \rho _{\rm n}}
\right) _{T _{\rm n}} , ~~
\Lambda _{\rm n T} \equiv 
\left( \frac{\partial \Lambda _{\rm n}}{\partial T_{\rm n}}
\right) _{\rho _{\rm n}} , ~~
\Lambda _{\rm i \rho} \equiv 
\left( \frac{\partial \Lambda _{\rm i}}{\partial \rho _{\rm i}}
\right) _{T _{\rm i}} , ~~
\Lambda _{\rm i T} \equiv 
\left( \frac{\partial \Lambda _{\rm i}}{\partial T_{\rm i}}
\right) _{\rho _{\rm i}} ,
\end{displaymath}
and
\begin{displaymath}
\nu_{\rm ni 0} \equiv \gamma _\nu \rho_{\rm i 0}
.
\end{displaymath}

We assume that the fluid is weakly ionized (i.e., $\rho _{{\rm i}0} \ll \rho _{{\rm n}0}$)
 and that temperatures of neutral and ion components are the same in the equilibrium (i.e., $T_{{\rm n}0} = T_{{\rm i}0}$).
According to $p={\cal R} \rho T / \mu$, assumptions $\rho _{{\rm i}0} \ll \rho _{{\rm n}0}$ and $T_{{\rm n}0} = T_{{\rm i}0}$ lead to 
$p _{{\rm i}0} \ll p _{{\rm n}0}$.
Using these approximations and the definition $\chi \equiv {\rho _{{\rm i}0}} / {\rho _{{\rm n} 0} }$, 
we can replace several expressions as follows:
\begin{eqnarray}
\frac{\rho _{{\rm i} 0} + \rho _{{\rm n} 0}}{\rho _{{\rm i} 0}}
\thicksim
\frac{1}{\chi},
\nonumber \\
\frac{p_{{\rm n} 0}} { \rho _{{\rm i} 0} + \rho _{{\rm n} 0} }
\thicksim
\frac{v _{\rm sn}^2}{\gamma},
\nonumber \\
\frac{p_{{\rm i} 0}} { \rho _{{\rm i} 0} + \rho _{{\rm n} 0} }
\thicksim
\frac{v _{\rm si}^2 \chi}{\gamma},
\nonumber \\
\gamma \frac{p_{{\rm i} 0} + p_{{\rm n} 0}}{\rho _{{\rm i} 0} + \rho _{{\rm n} 0}}
\thicksim
v _{\rm sn}^2,
\nonumber \\
\frac{p_{{\rm n} 0} + p_{{\rm i} 0} \gamma}{\rho _{{\rm i} 0} + \rho _{{\rm n} 0}}
\thicksim
\frac{v _{\rm sn}^2}{\gamma},
\nonumber \\
\frac{p_{{\rm i} 0} + p_{{\rm n} 0} \gamma}{\rho _{{\rm i} 0} + \rho _{{\rm n} 0}}
\thicksim
v _{\rm sn}^2,
\nonumber 
\end{eqnarray}
and
\begin{eqnarray}
\frac{B_{0}^2}{4 \pi} \frac{1}{\rho _{{\rm i} 0} + \rho _{{\rm n} 0}}
\thicksim
v _{\rm A}^2 \chi.
\nonumber
\end{eqnarray}
Then, the dispersion relation (\ref{dis-rela_partial}) is approximated to be 
\clearpage
\begin{eqnarray}
\lefteqn{
\left[
n^3 
+ n^2 \left(v_{\rm sn} k_{\rm nT} + v_{\rm sn} k \frac{k}{k_{\rm n K}} \right)
+ n (v_{\rm sn} k)^2 
+ \frac{1}{\gamma} (v_{\rm sn} k)^2
\left( v_{\rm sn} k_{\rm nT} - v_{\rm sn} k_{\rm n \rho}
+ v_{\rm sn} k \frac{k}{k_{\rm n K}} \right)  
\right] 
}
\nonumber \\
& & {}
\lefteqn{
\times 
\left[
n^3 
+ n^2 \left(v_{\rm si} k_{\rm iT} + v_{\rm si} k \frac{k}{k_{\rm i K}} \right)
+ n (v_{\rm si} k)^2 
+ n (v_{\rm A} k)^2
\right.
}
\nonumber \\
& & {}
\hspace{1.1cm}
\lefteqn{\left.
+ \frac{1}{\gamma}
\left( v_{\rm si} k_{\rm iT} + v_{\rm si} k \frac{k}{k_{\rm i K}} \right)
\left( v_{\rm si}^2 k^2 + \gamma v_{\rm A}^2 k^2 \right)
- \frac{1}{\gamma} (v_{\rm si} k)^2 (v_{\rm si} k_{\rm i \rho})
\right]
}
\nonumber \\
& = &
-\nu _{{\rm ni} 0} \frac{1}{\chi} \cdot
\left \{
n^5
+ n^4 
\left( 
v_{\rm si} k_{\rm iT} + v_{\rm si} k \frac{k}{k_{\rm i K}} 
+ v_{\rm sn} k_{\rm nT} + v_{\rm sn} k \frac{k}{k_{\rm n K}} 
\right)
\right.
\nonumber \\
& & {}
+ n^3 
\left[ 
v_{\rm sn} k \frac{k}{k_{\rm nK}} 
v_{\rm si} k \frac{k}{k_{\rm iK}}
+ v_{\rm sn} k \frac{k}{k_{\rm nK}} v_{\rm si} k_{\rm iT}
+ v_{\rm si} k \frac{k}{k_{\rm iK}} v_{\rm sn} k_{\rm nT}
\right.
\nonumber \\
& & {}
\hspace{1.1cm}
\left.
+ v_{\rm sn} k_{\rm nT} v_{\rm si} k_{\rm iT}
+ (v_{\rm sn} k)^2 + (v_{\rm A} k)^2 \chi
\right]
\nonumber \\
& & {}
+ n^2 
\left[
\frac{1}{\gamma}
(v_{\rm sn} k)^2 
(v_{\rm sn} k_{\rm nT} + v_{\rm sn} k \frac{k}{k_{\rm n K}})
- \frac{1}{\gamma} (v_{\rm sn} k)^2 v_{\rm sn} k_{\rm n \rho}
\right.
\nonumber \\
& & {}
\hspace{1.1cm}
\left.
+ (v_{\rm sn} k)^2 
(v_{\rm si} k_{\rm iT} + v_{\rm si} k \frac{k}{k_{\rm i K}})
- \frac{1}{\gamma} \chi (v_{\rm si} k)^2 v_{\rm si} k_{\rm i \rho}
\right.
\nonumber \\
& & {}
\hspace{1.1cm}
\left.
+ (v_{\rm A} k)^2 \chi 
\left( 
v_{\rm si} k_{\rm iT} + v_{\rm si} k \frac{k}{k_{\rm i K}} 
+ v_{\rm sn} k_{\rm nT} + v_{\rm sn} k \frac{k}{k_{\rm n K}} 
\right)
\right]
\nonumber \\
& & {} 
+ n 
\left[
( v_{\rm A}^2 k^2 \chi + \frac{1}{\gamma} v_{\rm sn}^2 k^2 )
\right.
\nonumber \\
& & {}
\hspace{1.1cm}
\left.
\cdot
\left(
v_{\rm sn} k \frac{k}{k_{\rm n K}} v_{\rm si} k \frac{k}{k_{\rm i K}}
+ v_{\rm sn} k \frac{k}{k_{\rm n K}} v_{\rm si} k_{\rm iT} 
+ v_{\rm si} k \frac{k}{k_{\rm i K}} v_{\rm sn} k_{\rm nT}
+ v_{\rm sn} k_{\rm nT} v_{\rm si} k_{\rm iT}
\right)
\right.
\nonumber \\
& & {}
\hspace{1.1cm}
\left.
\left.
- \frac{1}{\gamma} \chi (v_{\rm si} k)^2 v_{\rm si} k_{\rm i \rho}
(v_{\rm sn} k_{\rm nT} + v_{\rm sn} k \frac{k}{k_{\rm n K}})
- \frac{1}{\gamma} (v_{\rm sn} k)^2 v_{\rm sn} k_{\rm n \rho}
( v_{\rm si} k_{\rm iT} + v_{\rm si} k \frac{k}{k_{\rm i K}} )
\right]
\right\}
.
\nonumber
\end{eqnarray}
This is the dispersion relation (\ref{dis-rela}).

\section{Properties of the Dispersion Relation}
\label{subsec:nature_dispersion_relation}

The properties of thermal instability of one fluid approximation 
is briefly mentioned.
According to \citet{F65}, the criterion of the condensation mode of the instability for the neutral component with or without the magnetic field is
\begin{equation}
\frac{\partial \Lambda _{\rm n}}{\partial T _{\rm n}} 
- \frac{\rho _{{\rm n}0}}{T _{{\rm n}0}} \frac{\partial \Lambda _{\rm n}}{\partial \rho _{\rm n}}
< - k^2 \frac{K _{{\rm n}0} }{\rho _{{\rm n}0}}.
\label{crite_cond}
\end{equation}
If the cooling becomes efficient with temperature diminishing, 
the system is thermally unstable. 
This is expressed at the first term on the left-hand side of equation (\ref{crite_cond}).
If the cooling becomes efficient according to the contraction, 
the system is also thermally unstable. 
This is expressed at the second term on the left-hand side.
The thermal conduction smoothes the perturbation to suppress the instability.
It is efficient at small-scale events. 
This is expressed at the term on the right-hand side.

Mathematically, whether the system is stable or not depends on 
whether the sign of the real part of the growth rate $n$ is positive or negative. 
The system is unstable when the sign of the real part of $n$ is positive.
The sign of the real part of the solution of the dispersion relation is determined via algebra.

Similarly, the criterion of the condensation mode of 
the instability of the ion in the magnetic field is
\begin{equation}
\frac{\partial \Lambda _{\rm i}}{\partial T _{\rm i}} 
\left(1 + \gamma \frac{v _{\rm A}^2}{v _{\rm si}^2} \right)
- \frac{\rho _{{\rm i}0}}{T _{{\rm i}0}} \frac{\partial \Lambda _{\rm i}}{\partial \rho _{\rm i}}
< - k^2 \frac{K _{{\rm i} 0}}{\rho _{{\rm i}0}}
\left(1 + \gamma \frac{v _{\rm A}^2}{v _{\rm si}^2} \right).
\label{crite_ion}
\end{equation}
If
$ {\partial \Lambda _{\rm i}}/{\partial T _{\rm i}} > 0$, 
the magnetic field always suppresses the instability.
When
$ {\partial \Lambda _{\rm i}}/{\partial T _{\rm i}} < 0$, 
the criterion (\ref{crite_ion}) is rewritten as 
\begin{equation}
- \frac{\rho _{{\rm i}0}}{T _{{\rm i}0}} \frac{\partial \Lambda _{\rm i}}{\partial \rho _{\rm i}}
< - \left( \frac{\partial \Lambda _{\rm i}}{\partial T _{\rm i}}
+ k^2 \frac{K _{{\rm i} 0}}{\rho _{{\rm i}0}} \right)
\left(1 + \gamma \frac{v _{\rm A}^2}{v _{\rm si}^2} \right).
\end{equation}
Then, 
the effect of the magnetic field on the instability depends on the sign of 
${\partial \Lambda _{\rm i}} / {\partial T _{\rm i}}
+ k^2 {K _{{\rm i} 0}} / {\rho _{{\rm i}0}}$.
We note that this depends on the wave number $k$ .

The critical wavelength, i.e., $\lambda _{\rm F}$, 
above which the system is thermally unstable, is derived
 from equations (\ref{crite_cond}) for the neutral, or (\ref{crite_ion}) for the ion, respectively. 
The critical wavelength for the ion component depends on the magnetic field, while that for the neutral does not.
The criterion (\ref{crite_cond}) is rewritten as 
\begin{equation}
\frac{\lambda _{\rm F}}{2 \pi}
=
\frac{1}{k} 
=
\sqrt{ \frac{K _{{\rm n}0}}{\rho _{{\rm n}0}} }
\cdot
\left(
\frac{\rho _{{\rm n}0}}{T _{{\rm n}0}} \frac{\partial \Lambda _{\rm n}}{\partial \rho _{\rm n}}
- \frac{\partial \Lambda _{\rm n}}{\partial T _{\rm n}} 
\right) ^{-\frac{1}{2}}
,
\label{eq:appendix:criticallength:neutral}
\end{equation}
the criterion (\ref{crite_ion}) is rewritten as 
\begin{equation}
\frac{\lambda _{\rm F}}{2 \pi}
=
\frac{1}{k} 
=
\sqrt{ \frac{K _{{\rm i} 0}}{\rho _{{\rm i}0}} 
\cdot
\left( 1 + \gamma \frac{v _{\rm A}^2}{v _{\rm si}^2} 
\right)
}
\times
\left[
\frac{\rho _{{\rm i}0}}{T _{{\rm i}0}} \frac{\partial \Lambda _{\rm i}}{\partial \rho _{\rm i}}
- \frac{\partial \Lambda _{\rm i}}{\partial T _{\rm i}} 
\left(1 + \gamma \frac{v _{\rm A}^2}{v _{\rm si}^2} \right)
\right] ^{-\frac{1}{2}}
,
\end{equation}
respectively.

Moreover, the effect of the thermal conduction is mentioned. 
First, 
Observing to the conduction term 
$\nabla (K_{\rm n} \nabla T_{\rm n})$ in equation (\ref{eq:energy_n}) 
or the right-hand side of the criterion (\ref{crite_cond}), 
the conduction is more effective inversely proportionally 
to square of $\lambda$. 
Then, the conduction becomes less effective at large scale.
Second, 
according to the criterion (\ref{crite_cond}), 
the instability of the condensation mode prefers 
\begin{displaymath}
\frac{\rho _{{\rm n}0}}{T _{{\rm n}0}} \frac{\partial \Lambda _{\rm n}}{\partial \rho _{\rm n}}
\gg
\frac{\partial \Lambda _{\rm n}}{\partial T _{\rm n}}
.
\end{displaymath}
The preference allows us to approximate
\begin{eqnarray}
\lambda_{\rm F}
\nonumber
& \sim &
\sqrt{
\frac{K _{{\rm n}0}}{\rho _{{\rm n}0}} \cdot
T _{{\rm n}0}
\left[
\rho _{{\rm n}0} \frac{\partial \Lambda _{\rm n}}{\partial \rho _{\rm n}}  
\right] ^{-1}
} 
\\ \nonumber
& \sim & {}
\sqrt{
\frac{K _{{\rm n}0}}{\rho _{{\rm n}0}} \cdot
\frac{T _{{\rm n}0}}{\Lambda _{\rm n}}
}
\\ 
& = & {}
\lambda _{\rm F (e)}
\sqrt{
\frac{K _{{\rm n}0} T _{{\rm n}0}} {(\lambda _{\rm F (e)}) ^2} \cdot
\left(
\rho _{{\rm n}0}  \Lambda _{\rm n}
\right)^{-1}
}
.
\label{eq:lambdaF_rela_thermalconduction_sankou001}
\end{eqnarray}
According to equation (\ref{eq:energy_n}), 
$K _{{\rm n}0} T _{{\rm n}0} / (\lambda _{\rm F (e)})^2$ corresponds to 
the thermal conduction and $\rho _{{\rm n}0} \Lambda _{\rm n}$ to 
the cooling rate.
This equation (\ref{eq:lambdaF_rela_thermalconduction_sankou001}) 
indicates that the critical wavelength $\lambda _{\rm F (e)}$ is determined at 
the point where the effect of the thermal 
conduction is equivalent to that of the cooling.
The larger the thermal conduction is, 
the more the system stabilizes especially at smaller scale, 
and then the critical wavelength is enlarged.

\section{Normalization of the Dispersion Relation}
\label{Ap_nor}

We define the following quantities:
\begin{eqnarray}
Y \equiv \frac{n}{k_{{\rm n} \rho} v_{\rm sn}}
,~
X \equiv \frac{k}{k_{{\rm n} \rho}}
,~
C_{\nu} \equiv \frac{\nu_{{\rm ni}0}}{k_{{\rm n} \rho} v_{\rm sn}}
,
\nonumber
\end{eqnarray}
\begin{eqnarray}
\alpha _{\rm n} \equiv \frac{k_{{\rm n} T}}{k_{{\rm n} \rho}}
,~
\beta _{\rm n} \equiv \frac{k_{{\rm n} \rho}}{k_{{\rm n} K}}
,~
\alpha _{\rm i} \equiv \frac{k_{{\rm i} T}}{k_{{\rm i} \rho}}
,~
\beta _{\rm i} \equiv \frac{k_{{\rm i} \rho}}{k_{{\rm i} K}}
,~
\kappa _{v _{\rm A}} \equiv \frac{v_{\rm A}}{v _{\rm sn}}
\nonumber
,
\end{eqnarray}
and
\begin{eqnarray}
\kappa _{\rho} \equiv \frac{k_{{\rm i} \rho}}{k_{{\rm n} \rho}}
.
\end{eqnarray}
According to $v_{\rm s}^2 = \gamma ({\cal R} / {\mu}) T$, 
\begin{equation}
\frac{v_{\rm si}}{v _{\rm sn}} 
= 
\sqrt{ \frac{\mu _{\rm n} T_{{\rm i}0}}{\mu _{\rm i} T_{{\rm n}0}} }.
\end{equation}
Assuming $T_{{\rm i} 0} \sim T_{{\rm n} 0}$, 
\begin{equation}
\frac{v_{\rm si}}{v _{\rm sn}} 
= 
\sqrt{ \frac{\mu _{\rm n} T_{{\rm i}0}}{\mu _{\rm i} T_{{\rm n}0}} }
\sim
\sqrt{ \frac{\mu _{\rm n}}{\mu _{\rm i}} }
\equiv  \mu _{\rm ni} .
\end{equation}
Using these quantities, the dispersion relation (\ref{dis-rela}) is normalized as follows:
\begin{eqnarray}
\lefteqn{
\left\{
Y^3 
+ Y^2 ( \alpha _{\rm n} + X^2 \beta _{\rm n})
+ Y X^2
+ \frac{X^2}{\gamma} (\alpha _{\rm n} + X^2 \beta _{\rm n} - 1)
\right\}
}
\nonumber \\
& & {}
\times
\left\{
Y^3 
+ Y^2 \left( \mu_{\rm ni} \kappa _{\rho} \alpha _{\rm i} 
+ \mu_{\rm ni} \frac{\beta _{\rm i}}{\kappa _{\rho}} X^2 \right)
+ Y (\mu_{\rm ni} X)^2 
+ Y(\kappa _{v _{\rm A}} X)^2
\right.
\nonumber \\
& & {}
\hspace{0.8cm}
\left.
+ \frac{1}{\gamma} \left( \mu_{\rm ni} \kappa _{\rho} \alpha _{\rm i} 
+ \mu_{\rm ni} \frac{\beta _{\rm i}}{\kappa _{\rho}} X^2 \right)
[( \mu_{\rm ni} X)^2 + \gamma (\kappa _{v _{\rm A}} X)^2]
- \frac{1}{\gamma} (\mu_{\rm ni} X)^2 \mu_{\rm ni} \kappa _{\rho}
\right\}
\nonumber \\ 
& = & - \frac{C_{\nu}}{\chi}
\left\{
Y^5
+ Y^4 \left[\mu_{\rm ni} \kappa _{\rho} \alpha _{\rm i}
+ \mu_{\rm ni} \frac{\beta _{\rm i}}{\kappa _{\rho}} X^2
+ \alpha _{\rm n}
+ X^2 \beta _{\rm n} \right]
\right.
\nonumber \\ 
& & {}
\left.
+ Y^3 \left[
(\alpha _{\rm n} + X^2 \beta _{\rm n})
\left( \mu_{\rm ni} \frac{\beta _{\rm i}}{\kappa _{\rho}} X^2
+ \mu_{\rm ni} \kappa _{\rho} \alpha _{\rm i} \right)
+ X^2
+ (\kappa _{v _{\rm A}} X)^2 \chi 
\right]
\right.
\nonumber \\ 
& & {}
\left.
+ Y^2 \left[
\frac{1}{\gamma} X^2 (\alpha _{\rm n} + X^2 \beta _{\rm n} - 1)
+ X^2 \left( \mu_{\rm ni} \kappa _{\rho} \alpha _{\rm i}
+ \mu_{\rm ni} \frac{\beta _{\rm i}}{\kappa _{\rho}} X^2 \right)
- \frac{\chi}{\gamma}(\mu_{\rm ni} X)^2 \mu_{\rm ni} \kappa _{\rho}
\right.
\right.
\nonumber \\ 
& & {}
\hspace{1.1cm}
\left.
\left.
+ (\kappa _{v _{\rm A}} X)^2 \chi 
\left( \alpha _{\rm n} + X^2 \beta _{\rm n}
+\mu_{\rm ni} \kappa _{\rho} \alpha _{\rm i}
+ \mu_{\rm ni} \frac{\beta _{\rm i}}{\kappa _{\rho}} X^2 \right)
\right]
\right.
\nonumber \\ 
& & {}
\left.
+ Y \left[
\left(\kappa _{v _{\rm A}}^2 X^2 \chi + \frac{X^2}{\gamma} \right)
(\alpha _{\rm n} + X^2 \beta _{\rm n})
\left(\mu_{\rm ni} \frac{\beta _{\rm i}}{\kappa _{\rho}} X^2
+ \mu_{\rm ni} \kappa _{\rho} \alpha _{\rm i}
\right)
\right.
\right.
\nonumber \\ 
& & {}
\hspace{1.1cm}
\left.
\left.
- \frac{1}{\gamma} \chi (\mu_{\rm ni} X)^2
\mu_{\rm ni} \kappa _{\rho} (\alpha _{\rm n} + X^2 \beta _{\rm n})
- \frac{1}{\gamma} X^2 \left( \mu_{\rm ni} \kappa _{\rho} \alpha _{\rm i}
+ \mu_{\rm ni} \frac{\beta _{\rm i}}{\kappa _{\rho}} X^2 \right)
\right]
\right\}
\label{norm_dis}
\end{eqnarray}
We numerically solve this normalized dispersion relation (\ref{norm_dis})
in the main text.




\clearpage

\clearpage

{\plottwo{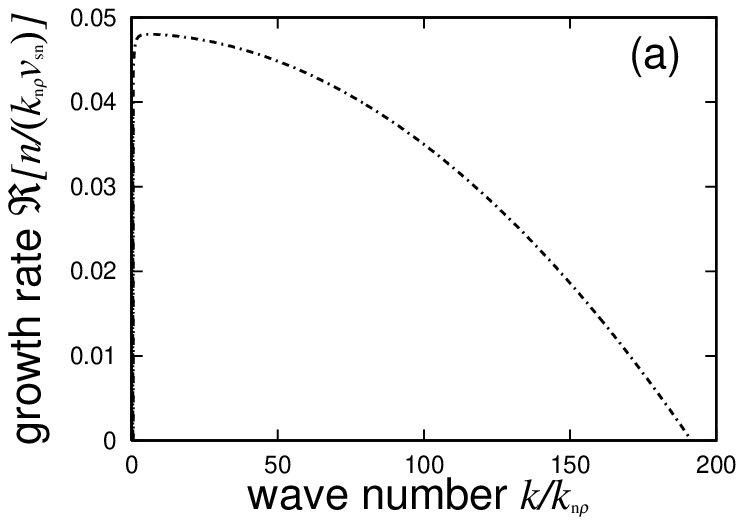}{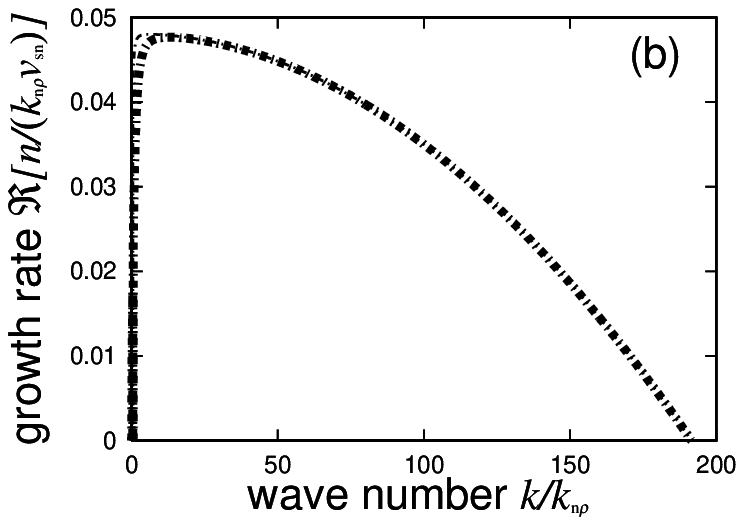}}
{\plottwo{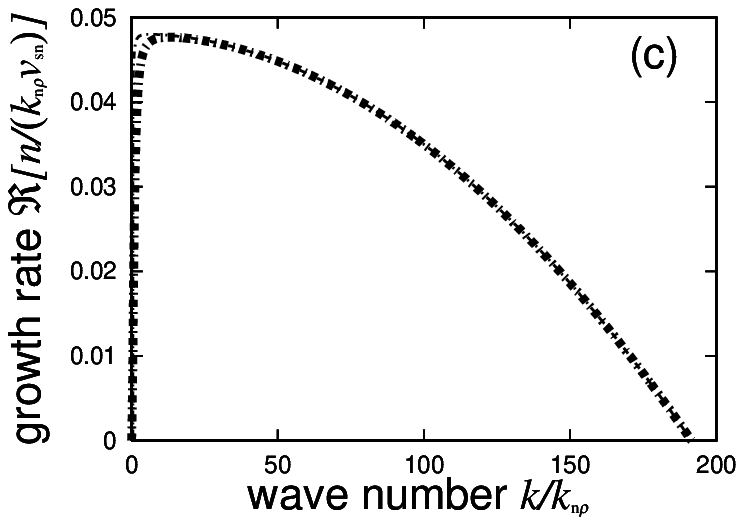}{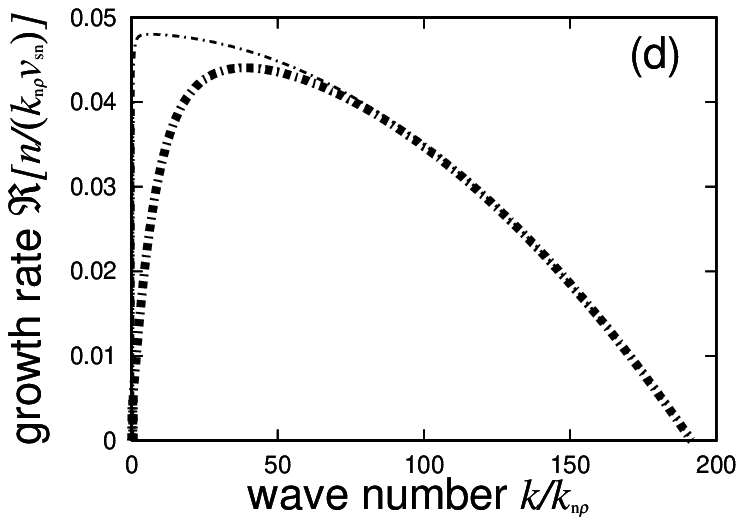}}
\epsscale{0.45}
\centerline{\plotone{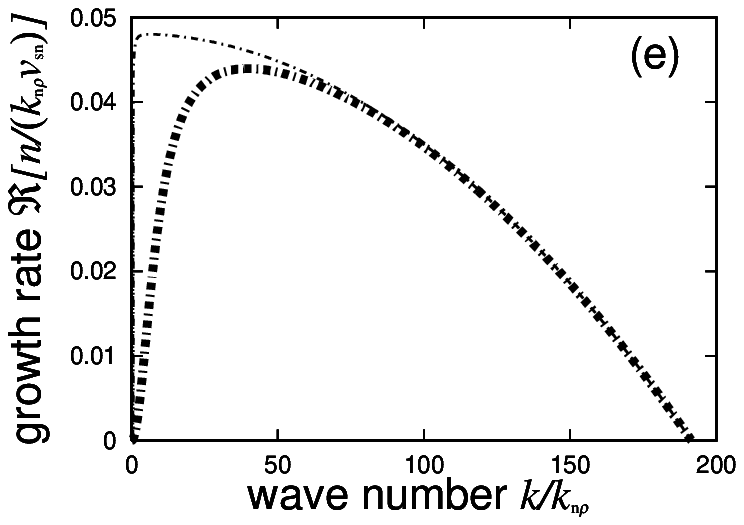}}
%
\clearpage
\begin{figure}
\caption
	{
	Growth rates of the condensation mode for the instability in the typical H {\rm \textsc{i}} region,
	where $T=100$ K and $n _{\rm neutral} = 71.9$ cm$^{-3}$. 
	These panels show the case when $f_{\rm CII} = 10^{-2}$ and $\chi = 10^{-6}$. 
    The other parameters are assumed as 
    $\langle \sigma v \rangle = 2 \times 10^{-9}$ cm$^3$ s$^{-1}$, 
    $K = 4.9 \times 10^3$ erg cm$^{-1}$ s$^{-1}$ K$^{-1}$,
    $A_{\rm C} = 3\times 10^{-4}$,
    $\mu_{n}=1$, $\mu_{i}=1/2$,
    $\gamma = 5/3$. 
    These panels are different in the strength of the magnetic field $B$ and the friction.
    (a) both of $B$ and the friction are zero; 
    (b) both of $B$ and the friction are the assuming typical values, i.e., $B = 10^{-6}$ Gauss (the thick dot-dashed curve); 
    (c) $B$ is hundred times as the assuming typical value, while the friction the assuming typical value (the thick dot-dashed curve); 
    (d) $B$ is the typical value, while the friction is hundred times as the assuming typical value (the thick dot-dashed curve); 
    (e) both of $B$ and the friction are hundred times as the assuming typical values (the thick dot-dashed curve). 
    For comparison, there are also plotted
     the condensation mode when $ B=0$ and $\nu _{{\rm ni}0} = 0$ (the thin dot-dashed curve) in each panel but (a).
	The horizontal axis is the normalized wave number $ k / k_{\rm n \rho} $. 
	The vertical axis corresponds to the normalized growth rate 
    $ \Re [n / ( k_{\rm n \rho} v_{\rm sn} ) ]$. 
	It is evaluated that $k_{\rm n \rho}= 2.43$ pc$^{-1}$ and 
    $k_{\rm n \rho} v_{\rm sn} = 9.28 \times 10^{-14}$ s.
    %
	It is noted that 
	$\alpha _{\rm n}=0.920$, $\alpha _{\rm i}=920000$, 
	$\beta _{\rm n}=2.19 \cdot 10^{-6}$, 
	$\beta _{\rm i}=2.73 \cdot 10^{-7}$, 
	$\kappa _{v _{\rm A}}=219$, $C_{\nu}=0.775$, 
	$\kappa _{\rho}= \chi / \mu_{\rm ni}^2 = 5 \cdot 10^{-7}$, 
	according to our notation summarized in Appendix \ref{Ap_nor}.
    }
\label{100K_kai10_-6_f_10-2_1}
\end{figure}
\clearpage


\epsscale{1.0}
{\plottwo{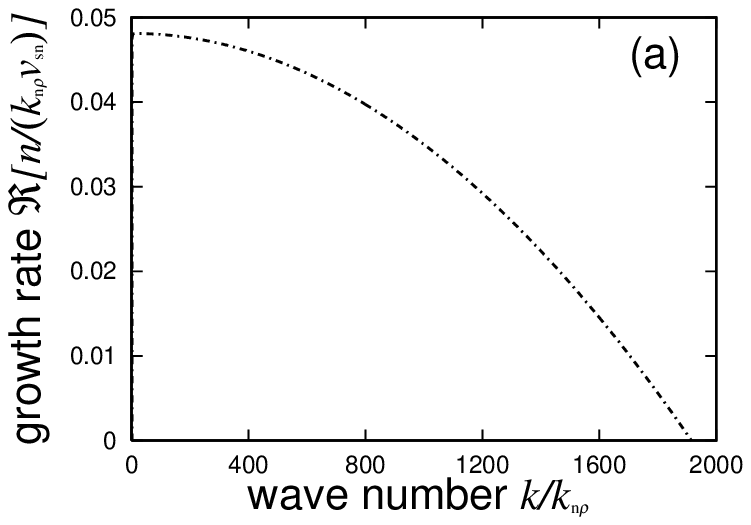}{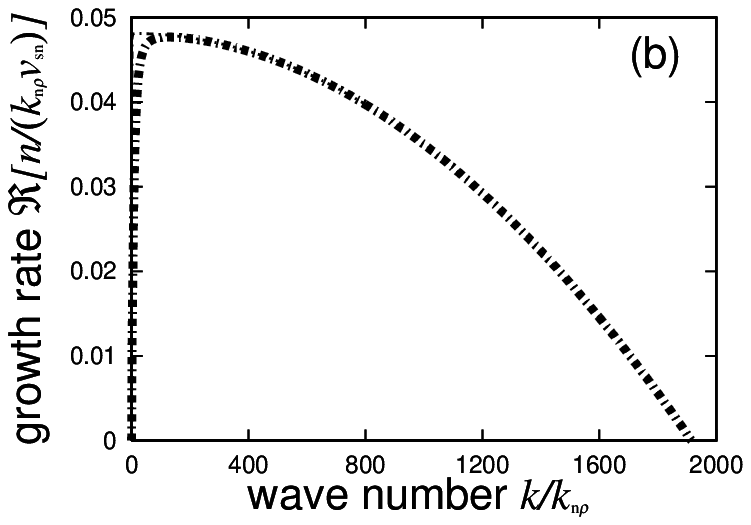}}
{\plottwo{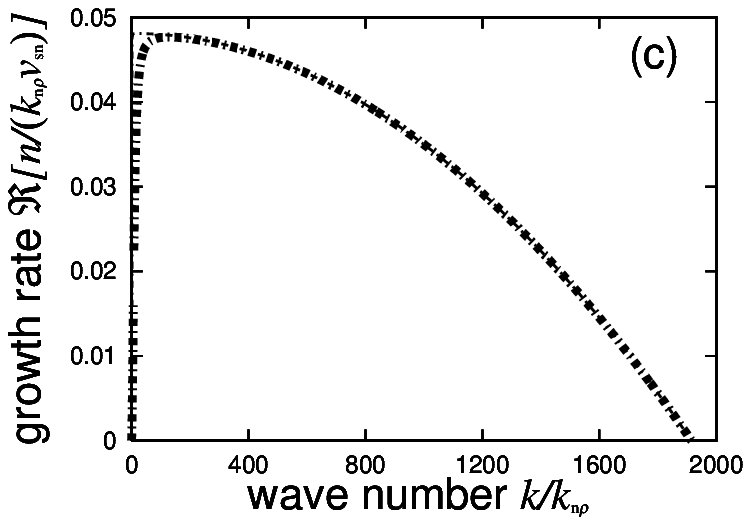}{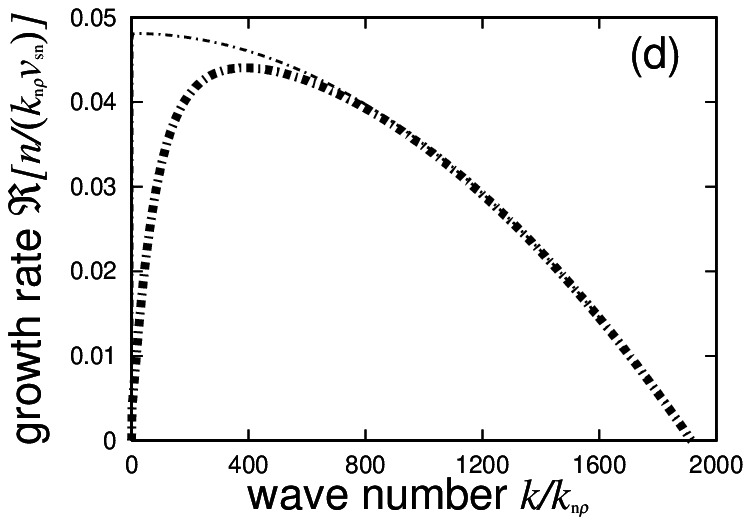}}
\epsscale{0.45}
\centerline{\plotone{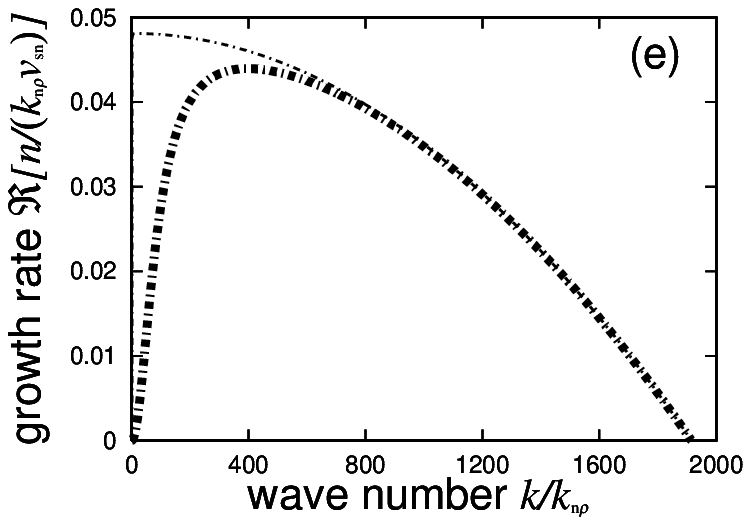}}
%
\clearpage
\begin{figure}
\caption
	{
	Growth rates of the condensation mode for the instability in the typical H {\rm \textsc{i}} region,
	where $T=100$ K and $n _{\rm neutral} = 71.9$ cm$^{-3}$. 
	These panels show the case when $f_{\rm CII} = 10^{-4}$ and $\chi = 10^{-6}$. 
    The other parameters are assumed as 
    $\langle \sigma v \rangle = 2 \times 10^{-9}$ cm$^3$ s$^{-1}$, 
    $K = 4.9 \times 10^3$ erg cm$^{-1}$ s$^{-1}$ K$^{-1}$,
    $A_{\rm C} = 3\times 10^{-4}$,
    $\mu_{n}=1$, $\mu_{i}=1/2$,
    $\gamma = 5/3$. 
    These panels are different in the strength of the magnetic field $B$ and the friction.
    (a) both of $B$ and the friction are zero; 
    (b) both $B$ and the friction is the assuming typical values, i.e., $B = 10^{-6}$ Gauss (the thick dot-dashed curve); 
    (c) $B$ is hundred times as the assuming typical value, while the friction the assuming typical value (the thick dot-dashed curve); 
    (d) $B$ is the typical value, while the friction is hundred times as the assuming typical value (the thick dot-dashed curve); 
    (e) both of $B$ and the friction are hundred times as the assuming typical values (the thick dot-dashed curve). 
    For comparison, there are also plotted
     the condensation mode when $ B=0$ and $\nu _{{\rm ni}0} = 0$ (the thin dot-dashed curve) in each panel but (a).
	The horizontal axis is the normalized wave number $ k / k_{\rm n \rho} $.
    The vertical axis corresponds to the normalized growth rate 
    $ \Re [n / ( k_{\rm n \rho} v_{\rm sn} ) ]$.
    It is evaluated that $k_{\rm n \rho}= 0.0243$ pc$^{-1}$ and 
    $k_{\rm n \rho} v_{\rm sn} = 9.28 \times 10^{-16}$ s.
    %
	Here, 
	$\alpha _{\rm n}=0.920$, $\alpha _{\rm i}=920000$, 
	$\beta _{\rm n}=2.19 \cdot 10^{-8}$, 
	$\beta _{\rm i}=2.73 \cdot 10^{-9}$, 
	$\kappa _{v _{\rm A}}=219$, $C_{\nu}=77.5$, 
	$\kappa _{\rho}= \chi / \mu_{\rm ni}^2 = 5 \cdot 10^{-7}$.
    }
\label{100K_kai10_-6_f_10-4_1}
\end{figure}
\clearpage


\epsscale{1.0}
{\plottwo{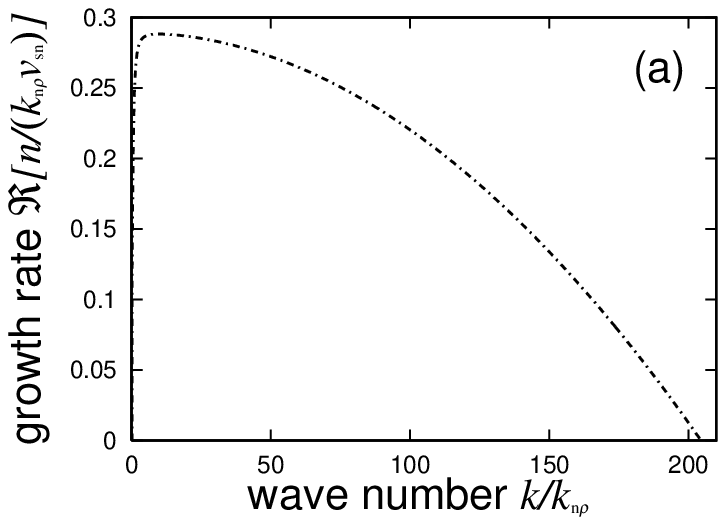}{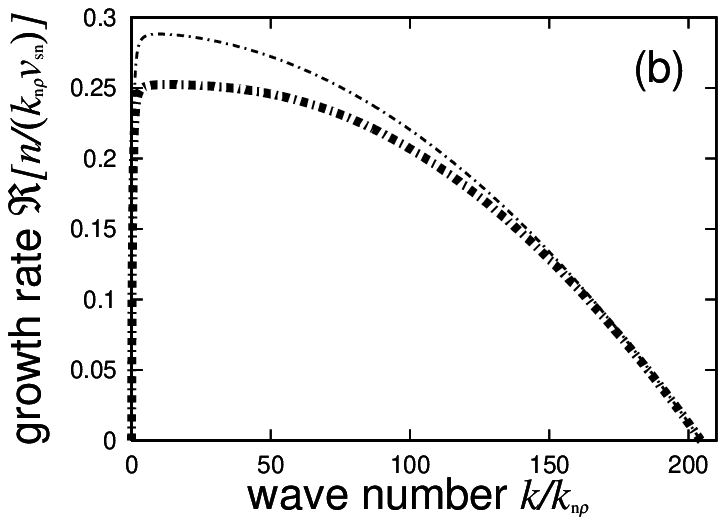}}
{\plottwo{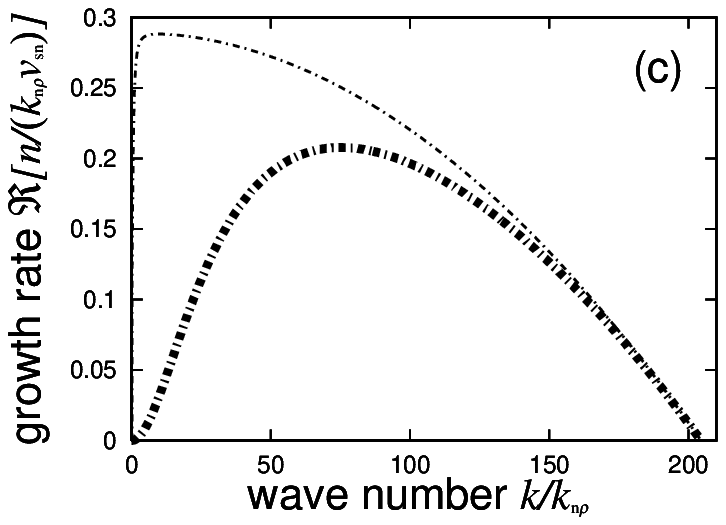}{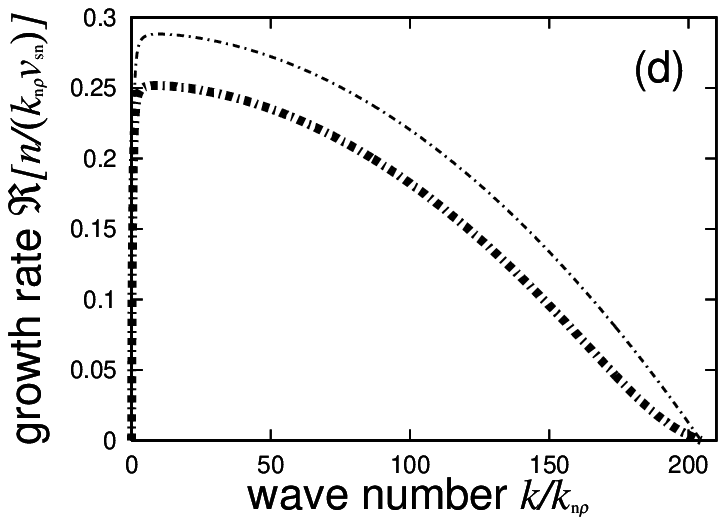}}
\epsscale{0.45}
\centerline{\plotone{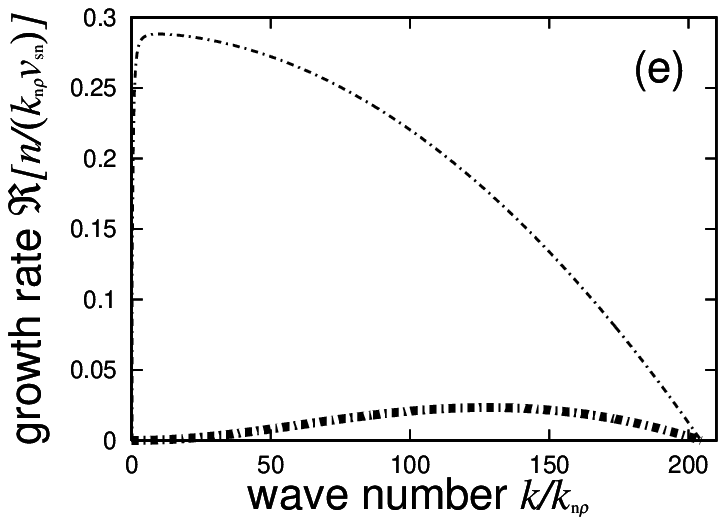}}
%
\clearpage
\begin{figure}
\caption
	{
	Growth rates of the condensation mode for the instability in the typical H {\rm \textsc{i}} region,
	where $T=100$ K and $n _{\rm neutral} = 71.9$ cm$^{-3}$. 
	These panels show the case when $f_{\rm CII} = 10^{-2}$ and $\chi = 10^{-2}$. 
    The other parameters are assumed as 
    $\langle \sigma v \rangle = 2 \times 10^{-9}$ cm$^3$ s$^{-1}$, 
    $K = 4.9 \times 10^3$ erg cm$^{-1}$ s$^{-1}$ K$^{-1}$,
    $A_{\rm C} = 3\times 10^{-4}$,
    $\mu_{n}=1$, $\mu_{i}=1/2$,
    $\gamma = 5/3$. 
    These panels are different in the strength of the magnetic field $B$ and the friction.
    (a) both of $B$ and the friction are zero; 
    (b) both of $B$ and the friction are the assuming typical values, i.e., $B = 10^{-6}$ Gauss (the thick dot-dashed curve); 
    (c) $B$ is hundred times as the assuming typical value, while the friction the assuming typical value (the thick dot-dashed curve); 
    (d) $B$ is the typical value, while the friction is hundred times as the assuming typical value (the thick dot-dashed curve); 
    (e) both of $B$ and the friction are hundred times as the assuming typical values (the thick dot-dashed curve). 
    This case is the extreme one.
    For comparison, there are also plotted
     the condensation mode when $ B=0$ and $\nu _{{\rm ni}0} = 0$ (the thin dot-dashed curve) in each panel but (a).
	The horizontal axis is the normalized wave number $ k / k_{\rm n \rho} $.
    The vertical axis corresponds to the normalized growth rate 
    $ \Re [n / ( k_{\rm n \rho} v_{\rm sn} ) ]$.
    It is evaluated that $k_{\rm n \rho}= 12.8 $ pc$^{-1}$ and 
    $k_{\rm n \rho} v_{\rm sn} = 4.90 \times 10^{-13}$ s.
	%
	We adopt 
	$\alpha _{\rm n}=0.517$, $\alpha _{\rm i}=51.7$, 
	$\beta _{\rm n}=1.15 \cdot 10^{-5}$, 
	$\beta _{\rm i}=1.44 \cdot 10^{-6}$, 
	$\kappa _{v _{\rm A}}=2.19$, $C_{\nu}=1470$, 
	$\kappa _{\rho}= \chi / \mu_{\rm ni}^2 = 5 \cdot 10^{-3}$.
    }
\label{100K_kai10_-2_f_10-2_1}
\end{figure}

\end{document}